\documentclass[12pt,reqno]{amsart}

\usepackage{fullpage}
\usepackage{amssymb, amsmath, amsxtra}
\usepackage{graphicx}
\usepackage{caption}
\usepackage{subcaption}
\usepackage{float}

\newtheorem{thm}{Theorem}[section]
\theoremstyle{plain}

\theoremstyle{plain} 

\theoremstyle{plain}

\theoremstyle{plain}

\newtheorem{prop}{Proposition}[section]

\def\X{{\rm X}}
\def\pr{{\rm pr}}

\def\sgn{{\rm sgn}}
\def\ns#1{\langle #1\rangle}
\def\Rnum{{\mathbb R}}

\def\parder#1{\partial_{#1}}

\def\const{\text{const.}}
\def\C{v}

\begin{document}
\allowdisplaybreaks[3]

\title{
A family of wave-breaking equations\\
generalizing the Camassa-Holm and Novikov equations
}

\author{
Stephen C. Anco$^1$, 
Priscila Leal da Silva$^2$, 
Igor Leite Freire$^2$
\\\lowercase{\scshape{
$^1$
Department of Mathematics and Statistics\\ 
Brock University\\
St. Catharines, Ontario, Canada, L2S 3A1\\
$^2$
Centro de Matem\'atica, Computa\c{c}\~ao e Cogni\c{c}\~ao\\ 
Universidade Federal do ABC - UFABC\\
Av. dos Estados, 5001, Bairro Bangu, 09210--580 
Santo Andr\'e, SP - Brazil\\
\rm e-mails: 
sanco@brocku.ca, 
priscila.silva@ufabc.edu.br,
igor.freire@ufabc.edu.br/igor.leite.freire@gmail.com
}}}

\begin{abstract}
A 4-parameter polynomial family of equations 
generalizing the Camassa-Holm and Novikov equations 
that describe breaking waves is introduced. 
A classification of low-order conservation laws, peaked travelling wave solutions, and Lie symmetries is presented for this family. 
These classifications pick out a 1-parameter equation that has several interesting features:
it reduces to the Camassa-Holm and Novikov equations 
when the polynomial has degree two and three; 
it has a conserved $H^1$ norm and it possesses $N$-peakon solutions, 
when the polynomial has any degree;
and it exhibits wave-breaking for certain solutions describing 
collisions between peakons and anti-peakons in the case $N=2$. 
\end{abstract}

\maketitle

\section{Introduction}
\label{intro}

There is considerable interest in the study of equations of the form 
$u_t -u_{txx} = f(u,u_x,u_{xx},u_{xxx})$ that describe breaking waves. 
In this paper we consider the equation 
\begin{equation}\label{4parmCHN}
u_t -u_{txx} + au^p u_x -b u^{p-1} u_x u_{xx} -c u^p u_{xxx}=0
\end{equation}
with parameters $a,b,c$ (not all zero) and $p\neq 0$. 
This 4-parameter family contains several integrable equations. 
For $(p,a,b,c)=(1,3,2,1)$ and $(p,a,b,c)=(1,4,3,1)$, 
equation \eqref{4parmCHN} reduces respectively to 
the Camassa-Holm equation \cite{CamHol}
\begin{equation}\label{CH}
u_t -u_{txx} + 3u u_x -2 u_x u_{xx} - u u_{xxx}=0
\end{equation}
and the Degasperis-Procesi equation \cite{DegPro}
\begin{equation}\label{D-P}
u_t -u_{txx} + 4u u_x -3 u_x u_{xx} - u u_{xxx}=0
\end{equation}
while for $(p,a,b,c)=(2,4,3,1)$, 
equation \eqref{4parmCHN} becomes the Novikov equation \cite{Nov}
\begin{equation}\label{N}
u_t -u_{txx} + 4u^2 u_x -3 u u_x u_{xx} -u^2 u_{xxx}=0 . 
\end{equation}
The three equations \eqref{CH}, \eqref{D-P}, \eqref{N}
are integrable in the sense of having 
a Lax pair, a bi-Hamiltonian structure, 
as well as hierarchies of local symmetries and local conservation laws,
and they also possess peaked travelling wave solutions. 

In addition to these integrable equations, 
many other non-integrable equations 
that admit breaking waves are included in the 4-parameter family \eqref{4parmCHN}. 
For instance, there is the $b$-equation 
\begin{equation}\label{CH-DP}
u_t -u_{txx} + (b+1)u u_x -b u_x u_{xx} - u u_{xxx}=0
\end{equation}
which unifies the Camassa-Holm and Degasperis-Procesi equations \cite{DegHolHon,HolHon}. 
There is also a modified version of the $b$-equation \cite{MiMu} 
\begin{equation}\label{mCH-DP}
u_t -u_{txx} + (b+1)u^2 u_x -b u u_x u_{xx} -u^2 u_{xxx}=0 
\end{equation}
which includes the Novikov equation. 
No other cases of the two equations \eqref{CH-DP} and \eqref{mCH-DP} 
are known to be integrable \cite{Nov,DegHolHon}. 

An equivalent form of the 4-parameter equation \eqref{4parmCHN} is given by 
\begin{equation}\label{4parmCHNm}
m_t + \tilde a u^p u_x +b u^{p-1} u_x m +c u^p m_x =0
\end{equation}
in terms of the momentum variable 
\begin{equation}\label{m}
m=u- u_{xx}
\end{equation}
with parameters 
\begin{equation}
\tilde a = a-b-c,
\quad
(\tilde a,b,c)\neq 0, 
\quad
p\neq 0 . 
\end{equation}
This parametric equation \eqref{4parmCHNm} is invariant under 
the group of scaling transformations
$u\rightarrow \lambda u$, $t\rightarrow \lambda^s t$, 
$(\tilde a,b,c) \rightarrow \lambda^{s+p} (\tilde a,b,c)$ with $\lambda\neq 0$. 

In section~\ref{conslaws}, 
we classify the low-order conservation laws of equation \eqref{4parmCHN}
and show that the Hamiltonians of the Camassa-Holm and Novikov equations 
are admitted as local conservation laws by equation \eqref{4parmCHN}
if and only if $\tilde a=0$ and $b=p+1$. 
We consider peaked travelling waves in section~\ref{peakons}
and use a weak formulation of equation \eqref{4parmCHN} to show that 
single peakon and multi-peakon solutions are admitted 
if and only if $\tilde a=0$ and $c\neq 0$ when $p\geq 0$. 
We derive the explicit equations of motion for $N\geq1$ peakon/anti-peakon solutions
and also obtain the constants of motion inherited from the local conservation laws of equation \eqref{4parmCHN}. 

In section~\ref{unifiedeqn}, 
we combine the previous results to obtain a natural 1-parameter family of equations 
\begin{equation}\label{CHNm}
m_t + (p+1) u^{p-1} u_x m + u^p m_x =0, 
\quad
p\geq 0 
\end{equation}
given by $\tilde a=0$, $b=p+1$, $c\neq 0$, $p\geq 0$,
where a scaling transformation $t\rightarrow t/c$
is used to put $c=1$. 
Since this 1-parameter family \eqref{CHNm} unifies the Camassa-Holm and Novikov equations,
we will refer to it as the {\em gCHN equation}. 
(Similar unified equations have been considered previously 
from related perspectives \cite{SilFre,Him1,Him2,AncRec}.)
We then discuss some general features of the dynamics of its $N\geq2$ peakon/anti-peakon solutions 
and we show that wave-breaking occurs for certain solutions describing 
collisions between peakons and anti-peakons in the case $N=2$. 

Finally, in section~\ref{remarks}, 
we make some concluding remarks including a possible scenario for wave-breaking in the Cauchy problem for weak solutions.

\section{Conservation laws}
\label{conslaws}

For the 4-parameter equation \eqref{4parmCHN}, 
a {\em local conservation law} \cite{Olv,2ndbook} is a space-time divergence 
\begin{equation}\label{conslaw}
D_t T +D_x X =0
\end{equation}
holding for all solutions $u(t,x)$ of equation \eqref{4parmCHN},
where the {\em conserved density} $T$ and the {\em spatial flux} $X$ 
are functions of $t$, $x$, $u$ and derivatives of $u$. 
The spatial integral of the conserved density $T$ satisfies
\begin{equation}
\frac d{dt} \int_{-\infty}^{\infty} T dx 
= -X\Big|_{-\infty}^{\infty}
\end{equation}
and so if the flux $X$ vanishes at spatial infinity, 
then 
\begin{equation}\label{C}
\mathcal C[u]= \int_{-\infty}^{\infty} T dx=\const
\end{equation}
formally yields a conserved quantity for equation \eqref{4parmCHN}.
Conversely, any such conserved quantity arises from
a local conservation law \eqref{conslaw}.

If the conserved quantity \eqref{C} is purely a boundary term, 
then the local conservation law is called {\em trivial}. 
This occurs when (and only when) 
the conserved density is a total $x$-derivative
and the flux is a total $t$-derivative, related by 
\begin{equation}\label{trivconslaw}
T = D_x\Theta, 
\quad
X = -D_t\Theta
\end{equation}
for all solutions $u(t,x)$ of equation \eqref{4parmCHN},
where $\Theta$ is some function of $t$, $x$, $u$ and derivatives of $u$. 
Two local conservation laws are \emph{equivalent} 
if they differ by a trivial conservation law, 
thereby giving the same conserved quantity up to boundary terms.

The set of all conservation laws (up to equivalence) 
admitted by equation \eqref{4parmCHN} 
forms a vector space on which there is a natural action 
\cite{Olv,2ndbook,BluTemAnc}
by the group of all Lie symmetries of the equation. 

For conserved densities and fluxes depending on at most 
$t,x,u,u_t,u_x,u_{tt},u_{tx},u_{xx}$, 
a conservation law can be expressed in an equivalent form 
by a divergence identity 
\begin{equation}\label{loworderchareqn}
D_t T +D_x X = 
(u_t -u_{txx} + au^p u_x -b u^{p-1} u_x u_{xx} -c u^p u_{xxx}) Q
\end{equation}
where
\begin{equation}\label{lowordermultiplier}
Q = -T_{u_{xx}} - X_{u_{tx}}
\end{equation}
is called the {\em multiplier}. 
This identity \eqref{loworderchareqn}--\eqref{lowordermultiplier} is called 
the {\em characteristic equation} \cite{Olv,2ndbook} 
for the conserved density and flux. 
By balancing the highest order $t$-derivative terms $u_{ttt}$ 
on both sides of the equation, 
we directly find that $T_{u_{tt}}=0$ and $X_{u_{tt}u_{tt}}=0$. 
Then balancing the terms $u_{tt}$, 
we see that $X_{u_{tt}u_{tx}}=0$. 
Hence the conserved density and the flux in the divergence identity 
must have the form 
\begin{equation}\label{loworderTX}
\begin{gathered}
T=T_0(t,x,u,u_t,u_x,u_{tx},u_{xx}), 
\\
X=X_0(t,x,u,u_t,u_x,u_{tx},u_{xx})+ u_{tt}X_1(t,x,u,u_t,u_x,u_{xx}). 
\end{gathered}
\end{equation}
Its multiplier \eqref{lowordermultiplier} thus has the form 
\begin{equation}\label{loworderQ}
Q = Q_0(t,x,u,u_t,u_x,u_{tx},u_{xx}). 
\end{equation}

In general, 
the differential order of a local conservation law is defined to be 
the smallest differential order among all equivalent conserved densities.
A local conservation law is said to be of {\em low order} 
if the differential orders of $T$ and $X$ are both strictly less than 
the differential order of the equation.

Consequently, conserved densities and fluxes of the form \eqref{loworderTX}
comprise all possible low-order conservation laws of equation \eqref{4parmCHN}. 
The problem of finding all low-order conservations then reduces to 
the simpler problem of finding all low-order multipliers \eqref{loworderQ}. 
Since equation \eqref{4parmCHN} is an evolution equation, 
it has no Lagrangian formulation in terms of the variable $u$. 
In this situation, the problem of finding multipliers can be understood as 
a kind of adjoint \cite{AncBlu97} of the problem of finding symmetries. 

An {\em infinitesimal symmetry} \cite{Olv,1stbook,2ndbook} 
of equation \eqref{4parmCHN} is a generator 
\begin{equation}\label{symmX}
\hat\X=P\parder{u}
\end{equation}
whose coefficient $P$ is given by a function of $t$, $x$, $u$ and derivatives of $u$, 
such that the prolonged generator satisfies the invariance condition 
\begin{equation}\label{symmdeteq}
\begin{aligned}
0 & 
= \pr\hat\X \big(u_t -u_{txx} + au^p u_x -b u^{p-1} u_x u_{xx} -c u^p u_{xxx}\big)
\\& 
= D_t P-D_tD_x^2 P + a D_x(u^p P) -b u^{p-1} D_x(u_x D_x P) -c u^p D_x^3 P
\\&\qquad 
-(b(p-1) u^{p-2} u_x u_{xx} + cp u^{p-1} u_{xxx}) P
\end{aligned}
\end{equation}
holding for all solutions $u(t,x)$ of equation \eqref{4parmCHN}. 
The Lie symmetry group of equation \eqref{4parmCHN} 
is generated by infinitesimal symmetries \eqref{symmX} 
with coefficients of the form 
\begin{equation}\label{contactsymmchar}
P(t,x,u,u_t,u_x) . 
\end{equation}
If $P$ is at most linear in $u_t$ and $u_x$, 
then the resulting generator \eqref{symmX} will yield a group of point transformations \cite{Olv,1stbook},
whereas if $P$ is nonlinear in $u_t$ or $u_x$, 
then a group of contact transformations \cite{Olv,1stbook} will be generated. 
Hence, all generators of Lie symmetries admitted by equation \eqref{4parmCHN} 
are determined by the solutions of condition \eqref{symmdeteq}
for $P(t,x,u,u_t,u_x)$. 
(It is straightforward to solve this determining equation by Maple 
to classify the Lie symmetry group of equation \eqref{4parmCHN},
as shown in the Appendix.)

The condition for determining all multipliers $Q(t,x,u,u_t,u_x,u_{tx},u_{xx})$ 
of low-order conservation laws \eqref{loworderQ} 
admitted by equation \eqref{4parmCHN} 
consists of 
\begin{equation}\label{Qdeteq}
E_u\big( (u_t -u_{txx} + au^p u_x -b u^{p-1} u_x u_{xx} -c u^p u_{xxx}) Q \big)
=0
\end{equation}
which arises from the property that the variational derivative (Euler operator)
\begin{equation}
E_u=
\parder{u}-D_x\parder{u_x} -D_t\parder{u_t} 
+D_x^2\parder{u_{xx}} +D_t^2\parder{u_{tt}} +D_xD_t\parder{u_{tx}} 
-\cdots
\end{equation}
annihilates an expression identically iff it is a space-time divergence
\cite{Olv,2ndbook}. 
This condition \eqref{Qdeteq} can be split 
with respect to $u_{txx}$ and $t,x$-derivatives of $u_{txx}$, 
yielding an equivalent overdetermined system of equations on $Q$. 
One equation in this system is given by 
the adjoint of the symmetry determining equation \eqref{symmdeteq},
\begin{equation}\label{adjsymmdeteq}
\begin{aligned}
0 & 
= -D_t Q +D_tD_x^2 Q - au^p D_x Q -b D_x(u_x D_x(u^{p-1} Q)) + c D_x^3(u^p Q)
\\&\quad 
-(b(p-1) u^{p-2} u_x u_{xx} +cp u^{p-1} u_{xxx}) Q
\end{aligned}
\end{equation}
holding for all solutions $u(t,x)$ of equation \eqref{4parmCHN}. 
Solutions $Q$ of this equation \eqref{adjsymmdeteq} 
are called {\em adjoint-symmetries} (or {\em cosymmetries})
\cite{AncBlu97,1stbook,AncBlu02a,AncBlu02b}. 
The remaining equations in the system comprise Helmholtz conditions 
which are necessary and sufficient for $Q$ to have the form \eqref{lowordermultiplier}.
As a consequence, 
multipliers \eqref{loworderQ} are simply 
adjoint-symmetries that have a certain variational form. 

For any solution \eqref{loworderQ} of the multiplier determining equation \eqref{Qdeteq}, 
a corresponding conserved density and flux of the form \eqref{loworderTX}
can be recovered 
either through integration \cite{2ndbook} 
of the characteristic equation \eqref{loworderchareqn},
which splits with respect to $u_{ttx}$, $u_{txx}$, $u_{xxx}$, $u_{tt}$
into a system of equations for $T$ and $X$, 
or through a homotopy integral formula
\cite{2ndbook,Olv,DecNiv,PooHer} ,
which expresses $T$ and $X$ directly in terms of
$(u_t -u_{txx} + au^p u_x -b u^{p-1} u_x u_{xx} -c u^p u_{xxx}) Q$. 
It is straightforward to show that $T$ and $X$ 
have the form \eqref{trivconslaw} of a trivial conservation law 
iff $Q=0$. 
Thus there is a one-to-one correspondence between equivalence classes of 
non-trivial low-order conservation laws \eqref{loworderTX}
and non-zero low-order multipliers \eqref{loworderQ}. 

\subsection{Classification results}

Both the Camassa-Holm equation \eqref{CH} and Novikov equation \eqref{N} 
possess low-order local conservations law given by the conserved densities
\cite{CamHol,HonWan,Len05}
\begin{gather}
T=mu = u^2 + u_x^2 +D_x(-uu_x), 
\label{T1-CH-N}
\\
T=m^q, 
\label{T2-CH-N}
\end{gather}
where $q=1/2$ and $q=2/3$, respectively, for the two equations. 
In addition, 
the Camassa-Holm equation \eqref{CH} itself is a low-order local conservation law
having the conserved density 
\begin{equation} 
T = m = u +D_x(-u_x). 
\label{T3-CH-N}
\end{equation} 
All of these conserved densities are related to Hamiltonian structures 
for the two equations \cite{CamHol,HonWan,Len05}. 
The corresponding multipliers are respectively given by 
\begin{align}
& Q=u ,
\label{Q1-CH-N}
\\
& Q=qm^{q-1}, 
\label{Q2-CH-N}
\\
& Q=1. 
\label{Q3-CH-N}
\end{align}

To look for conserved densities of the same form for equation \eqref{4parmCHN}, 
we now classify all multipliers up to 1st-order 
\begin{equation}\label{Q1storder}
Q=Q(t,x,u,u_x,u_t)
\end{equation}
as well as all 2nd-order multipliers with the specific form 
\begin{equation}\label{Qfunctm}
Q=Q(u,u_x,u_{xx}) . 
\end{equation}
In each case it is straightforward to solve the determining equation \eqref{Qdeteq}
by use of Maple (as shown in the Appendix), 
which leads to the following classification result.

\begin{prop}\label{class-loworderQ}
(i) 
Equation \eqref{4parmCHN} admits 0th-order multipliers 
only in the following cases:
\begin{align}
{\rm (a)}\qquad
\label{0thQ-a}
& Q =1 
\quad\text{ iff }\quad
p=1 
\quad\text{ or }\quad
b=pc
\\\nonumber\\
{\rm (b)}\qquad
\label{0thQ-b}
& Q = u 
\quad\text{ iff }\quad
b=(p+1)c
\\\nonumber\\
{\rm (c)}\qquad
\label{0thQ-c}
& Q =\exp(\pm \sqrt{a/c}\; x)
\quad\text{ iff }\quad
p=1, 
\quad
b=3c
\\\nonumber\\
{\rm (d)}\qquad
\label{0thQ-d}
& Q =f(t)\exp(\pm x)
\quad\text{ iff }\quad
p=1, 
\quad
a=c,
\quad
b=3c 
\\\nonumber\\
{\rm (e)}\qquad
\label{0thQ-e}
& Q = x - ctu
\quad\text{ iff }\quad
p=1, 
\quad
a=c,
\quad
b=2c 
\end{align}
(ii) For any $p\neq 0$ and any $(a,b,c)\neq 0$, 
equation \eqref{4parmCHN} admits no 1st-order multipliers. 
\newline
(iii) Equation \eqref{4parmCHN} admits 2nd-order multipliers of the form \eqref{Qfunctm} only in the following cases:
\begin{align}
{\rm (a)}\qquad
\label{2ndQ-1}
& Q =(u-u_{xx})^{q-1} 
\quad\text{ iff }\quad
q=pc/b \neq 1, 
\quad
a=b+c
\\\nonumber\\
{\rm (b)}\qquad
\label{2ndQ-2}
& Q= 2a u - (p+2)c u_{xx}
\quad\text{ iff }\quad
b = \tfrac{1}{2} pc,
\quad
c\neq 0,
\quad
p\neq -2
\end{align}
\end{prop}

In light of the adjoint relationship between multipliers and symmetries,  
the classification of 0th- and 1st- order multipliers 
in Proposition~\ref{class-loworderQ}
is a counterpart of the classification of Lie symmetries 
(cf.\ Proposition~\ref{class-pointsymm}). 

Next we obtain the corresponding conserved densities and fluxes 
for each multiplier \eqref{0thQ-a}--\eqref{2ndQ-1} by 
first splitting the characteristic equation \eqref{loworderchareqn}
with respect to $u_{ttx}$, $u_{txx}$, $u_{xxx}$, $u_{tt}$
where $T$ and $X$ have the form \eqref{loworderTX}, 
and then integrating the resulting system of equations. 
This yields the following low-order local conservation laws
for equation \eqref{4parmCHN}. 

\begin{thm}\label{class-loworderTX}
(i) The local conservation laws admitted by the wave-breaking equation \eqref{4parmCHN} 
with multipliers of at most 1st-order consist of 
three 0th-order conservation laws
\begin{align}
\label{0thQTX-a}
&\begin{aligned}
& 
T_{1} = u,
\quad
X_{1} = \frac{a}{p+1}u^{p+1} +\tfrac{1}{2} (pc-b) u_x^2 -cu^p u_{xx} +u_{tx}
\\
& \text{ iff }\quad
p=1
\quad\text{ or }\quad
b=pc
;
\end{aligned}
\\
\label{0thQTX-c}
&\begin{aligned}
& 
T_{2} = (c-a)e^{\pm\sqrt{a/c}\; x} u,
\quad
X_{2} = e^{\pm\sqrt{a/c}\; x}
\big( \pm \sqrt{ac} (u_t + cu u_x) -c u_{tx} -c^2(u_x^2 + u u_{xx}) \big)
\\
& \text{ iff }\quad
p=1, 
\quad 
b=3c 
\quad
(c\neq 0) 
;
\end{aligned}
\\
\label{0thQTX-d}
&\begin{aligned}
& 
T_{3} = 0, 
\quad
X_{3} = f(t) e^{\pm x}
\big( \pm (u_t + cu u_x) - u_{tx} -c(u_x^2 + u u_{xx}) \big)
\\
& \text{ iff }\quad
p=1, 
\quad 
a=c, 
\quad
b=3c 
;
\end{aligned}
\end{align}
and two 1st-order conservation laws
\begin{align}
\label{0thQTX-b}
& \begin{aligned}
& 
T_{4} = \tfrac{1}{2}(u^2 + u_x^2),
\quad
X_{4} = \big( \frac{a}{p+2}u-c u_{xx}\big) u^{p+1} -uu_{tx}
\\
& \text{ iff }\quad
b=(p+1)c 
;
\end{aligned}
\\
\label{0thQTX-e}
& \begin{aligned}
& 
\begin{aligned}
T_{5} = -\tfrac{1}{2}c t(u^2 + u_x^2) +xu,\\
\quad
\end{aligned}
\quad
\begin{aligned}
& X_{5} = (ct u-x)(u_{tx} +cuu_{xx}) +u_t\\
&\qquad\qquad 
-\tfrac{1}{3} c^2 tu^3 +\tfrac{1}{2}c x(u^2-u_x^2+2uu_x)
\end{aligned}
\\
& \text{ iff }\quad
p=1,
\quad
a=c,
\quad
b=2c.
\end{aligned}
\end{align}
(ii) 
The local conservation laws admitted by the wave-breaking equation \eqref{4parmCHN} 
with 2nd-order multipliers of the form \eqref{Qfunctm}
consist of two 2nd-order conservation laws
\begin{align}
\label{2ndQTX-1}
& \begin{aligned}
&
T_{6} = (u-u_{xx})^{pc/b}, 
\quad
X_{6}= c u^p (u-u_{xx})^{pc/b}, 
\\
& \text{ iff }\quad
a=b+c
\quad
(b\neq pc,
\quad
c\neq 0) ;
\end{aligned}
\\
\label{2ndQTX-2}
& \begin{aligned}
& 
\begin{aligned}
T_{7} = a u^2 + (a+b+c)u_x^2 +(b+c) u_{xx}^2,\\
\quad
\end{aligned}
\quad
\begin{aligned}
& X_{7} = \frac{2}{p+2}(au-(b+c)u_{xx})^2 u^p\\
&\qquad\qquad 
-2au u_{tx} -2(b+c)u_tu_x
\end{aligned}
\\
& \text{ iff }\quad
b = \tfrac{1}{2} pc
\quad
(c\neq 0). 
\end{aligned}
\end{align}
In these conservation laws \eqref{0thQTX-a}--\eqref{2ndQTX-2}, 
any terms of the form $q^{-1} u^q$ in the case $q=0$ should be replaced by $\ln|u|$. 
\end{thm}

These conservation laws yield the following conserved integrals. 
We start with the conservation laws at 0th order. 
From $T_{1}$, we have 
\begin{equation}\label{0thC-a}
\mathcal C_1= \int_{-\infty}^{\infty} u\; dx , 
\qquad
p=1
\quad\text{ or }\quad
b=pc
\end{equation}
which is the conserved mass for equation \eqref{4parmCHN}. 
The conserved integral arising from $T_{2}$ is a weighted mass, 
\begin{equation}\label{0thC-c}
\mathcal C_2= \int_{-\infty}^{\infty} e^{\pm\sqrt{a/c}\; x} u \; dx, 
\qquad
p=1, 
\quad 
b=3c \neq 0
\quad
(c\neq a) . 
\end{equation}
Interestingly, from $T_{3}$ we get a conserved integral which vanishes, 
but has a non-zero spatial flux. 
This type of conservation law arises because the multiplier \eqref{0thQ-d}
converts equation \eqref{4parmCHN} into the form of a total $x$-derivative. 

Next we look at the conservation laws at 1st order. 
From $T_{4}$, the $H^1$ norm of $u(t,x)$ is conserved, 
\begin{equation}\label{0thC-b}
\mathcal C_4= \int_{-\infty}^{\infty} u^2 + u_x^2 \; dx 
= \|u\|_{H^1}, 
\qquad
b=(p+1)c . 
\end{equation}
From $T_{5}$, we have
\begin{equation}\label{0thC-e}
\begin{aligned}
\mathcal C_5 & = t\int_{-\infty}^{\infty} u^2 + u_x^2\; dx - \frac{2}{c}\int_{-\infty}^{\infty} xu\; dx \\
& = t\|u\|_{H^1} - \frac{2}{c}\mathcal P(t), 
\qquad
p=1,
\quad
a=c,
\quad
b=2c\neq 0
\end{aligned}
\end{equation}
where $\mathcal P(t)= \int_{-\infty}^{\infty} xu\; dx$ is the center of mass of $u(t,x)$. 
Since $\mathcal C_5$ is conserved, it can be evaluated at $t=0$, 
which yields the relation 
$\mathcal P(t) = \mathcal P(0) + (c/2) t\|u\|_{H^1}$. 
This shows that the center of mass moves at a constant speed controlled by
the $H^1$ norm of $u(t,x)$. 

Finally, we consider the conservation laws at 2nd order. 
From $T_{6}$, we get 
\begin{equation}\label{2ndC-1}
\mathcal C_6= \int_{-\infty}^{\infty} (u-u_{xx})^q \; dx ,
\qquad
a=b+c,
\quad
q=pc/b \neq 1.
\end{equation}
This shows that the $L^{q}$ norm of $m=u-u_{xx}$ is conserved
if $m$ does not change sign or if $q$ is an even integer. 
The conserved integral arising from $T_{7}$ is 
a linear combination of the $L^2$ norms of $u,u_x,u_{xx}$
as given by 
\begin{equation}\label{2ndC-2}
\begin{aligned}
\mathcal C_7  & = \int_{-\infty}^{\infty} a u^2 + (a+b+c)u_x^2 +(b+c) u_{xx}^2\; dx\\
& = a\|u\|_{L^2} +(a+b+c)\|u_x\|_{L^2} + (b+c) \|u_{xx}\|_{L^2}, 
\qquad
b = \tfrac{1}{2} pc \neq 0. 
\end{aligned}
\end{equation}
This can be written alternatively as a weighted $H^2$ norm when $b+c\neq 0$. 
It is interesting to note that simultaneous conservation of both
the $H^1$ and the weighted $H^2$ norms requires the condition 
$b=(p+1)c=\tfrac{1}{2}pc$ which holds iff $p=-2$ and $b+c=0$, 
but in this case $\mathcal C_7 = a\mathcal C_4 = a\|u\|_{H^1}$.

\section{Peakon solutions}
\label{peakons}

Both the Camassa-Holm and Novikov equations possess 
peaked travelling wave solutions \cite{CamHol,HonWan}, called peakons, 
\begin{equation}\label{peakonwave}
u(t,x) = \C{}^q \exp(-|x-\C t|),
\quad
\C=\const
\end{equation}
where $q=1$ and $q=1/2$, respectively, for the two equations. 
Peakons have attracted much attention in the study of breaking wave equations. 

In general, on $-\infty<x<\infty$, 
a peakon is a weak travelling wave solution satisfying 
an integral (i.e.\ weak) formulation of a breaking wave equation. 
Such a formulation is essential for deriving multi-peakon solutions. 
However, single peakons can be derived directly from the travelling wave 
reduction of a breaking wave equation,
which will be the approach we use here. 

\subsection{Single peakon solution}

The manifest invariance of the 4-parameter equation \eqref{4parmCHN} 
under time-translation and space-translation symmetries 
implies the existence of travelling wave solutions
\begin{equation}\label{travwave}
u=\phi(z),
\quad
z=x-\C t,
\quad
\C=\const\neq 0
\end{equation}
where $\phi(z)$ satisfies the ODE 
\begin{equation}\label{travwaveODE}
-\C(\phi-\phi'')' +a\phi^p\phi' -b\phi^{p-1}\phi'\phi'' -c\phi^p\phi'''
=0 . 
\end{equation}

For the travelling wave ODE \eqref{travwaveODE}, 
an integral formulation is obtained through 
multiplying this ODE by a test function $\psi$ 
(which is smooth and has compact support) 
and integrating over $-\infty<z<\infty$, 
leaving at most first derivatives of $\phi$ in the integral,
which yields
\begin{equation}\label{weaktravwave}
\begin{aligned}
0= \int_{-\infty}^{+\infty} & \Big(
\C(\psi''-\psi)\phi' + (a\psi-c\psi'')\phi^p\phi'
+ \tfrac{1}{2}(b-3pc)\psi' \phi^{p-1} \phi'{}^2 
\\&\qquad
+ \tfrac{1}{2}(p-1)(b-pc)\psi \phi^{p-2} \phi'{}^3 
\Big) dz . 
\end{aligned}
\end{equation}
A weak solution of ODE \eqref{travwaveODE} is a function 
$\phi(z)$ that belongs to the Sobolev space $W^{1,3}_{\rm loc}(\Rnum)$ 
and that satisfies the integral equation \eqref{weaktravwave}
for all smooth test functions $\psi(z)$ with compact support on $\Rnum$. 

To proceed we substitute a peaked travelling wave expression 
\begin{equation}\label{phi-peakon}
\phi = \alpha e^{-|z|},
\quad
\alpha=\const
\end{equation}
into equation \eqref{weaktravwave}
and split up the integral into the intervals 
$(-\infty,0)$ and $(0,+\infty)$. 
The first term in equation \eqref{weaktravwave} yields, 
after integration by parts, 
\begin{equation}\label{term1}
\int_{-\infty}^{0} \C(\psi''-\psi)\phi' \; dz
+ \int_{0}^{+\infty} \C(\psi''-\psi)\phi' \; dz
= 2\alpha \C \psi'(0) . 
\end{equation}
Similarly, the second term in equation \eqref{weaktravwave} gives
\begin{equation}\label{term2}
\begin{aligned}
& \int_{-\infty}^{0} (a\psi-c\psi'')\phi^p\phi' dz
+ \int_{0}^{+\infty} (a\psi-c\psi'')\phi^p\phi' dz
\\&\qquad
= -2\alpha^{p+1}c \psi'(0) 
+ \alpha^{p+1} (-a+(p+1)^2 c) \int_{-\infty}^{+\infty} \sgn(z) \psi e^{-(p+1)|z|}\; dz
\end{aligned}
\end{equation}
provided $p+1>0$ so that the boundary terms at $z=\pm\infty$ vanish. 
The third and fourth terms in equation \eqref{weaktravwave} together yield
\begin{equation}\label{term3+4}
\begin{aligned}
& \int_{-\infty}^{0} \Big( \tfrac{1}{2}(p-1)(b-pc)\psi \phi^{p-2} \phi'{}^3 
+ \tfrac{1}{2}(b-3pc)\psi' \phi^{p-1} \phi'{}^2 \Big) dz
\\&\qquad
+ \int_{0}^{+\infty} \Big( \tfrac{1}{2}(p-1)(b-pc)\psi \phi^{p-2} \phi'{}^3 
+ \tfrac{1}{2}(b-3pc)\psi' \phi^{p-1} \phi'{}^2 \Big) dz
\\&\qquad
= \alpha^{p+1}(b-p(p+2)c) \int_{-\infty}^{+\infty} \sgn(z) \psi e^{-(p+1)|z|}\; dz . 
\end{aligned}
\end{equation}
When the terms \eqref{term1}--\eqref{term3+4} are combined, 
we find that equation \eqref{weaktravwave} reduces to 
\begin{equation}
0 = 2\alpha (\C-c\alpha^p) \psi'(0) + \alpha^{p+1}(b+c-a) \int_{-\infty}^{+\infty} \sgn(z) \psi e^{-(p+1)|z|}\; dz . 
\end{equation}
This equation is satisfied for all test functions $\psi$ iff 
\begin{equation}
a=b+c,
\quad
c\alpha^p=\C, 
\end{equation}
which determines the amplitude $\alpha$ in the peakon expression \eqref{phi-peakon}. 
Thus we obtain the following result. 

\begin{prop}\label{class-peakon}
The travelling wave equation \eqref{weaktravwave} admits a peakon solution 
only in the case 
\begin{equation}\label{peakonsoln}
\phi(z) = (\C/c)^{1/p} e^{-|z|},
\quad
a=b+c,
\quad
c\neq0,
\quad
p+1>0
\end{equation}
where $\C=\const$ is the wave speed. 
\end{prop}

The resulting peakon solution of equation \eqref{4parmCHN} is given by 
\begin{equation}\label{peakon}
u(t,x) = c^{-1/p}\C^{1/p} \exp(-|x-\C t |),
\quad
a=b+c .
\end{equation}
When the nonlinearity power $p$ is a positive integer, 
then the wave speed is necessarily positive, $\C>0$, if $p$ is even, 
as in the case ($p=2$) of the Novikov equation \eqref{N},
while if $p$ is odd, the wave speed can be either positive or negative, 
$\C\gtrless 0$, 
as in the case ($p=1$) of the Camassa-Holm equation \eqref{CH}. 

The peakon solution \eqref{peakon} satisfies equation \eqref{4parmCHN}
only in the sense of a weak solution. 
This means $u(t,x)$ is a distribution 
in $L^\infty_{\rm loc}(-T,T)$ with respect to $t\in(-T,T)$ for some $T>0$
and in $W^{1,3}_{\rm loc}(\Rnum)$ with respect to $x\in\Rnum$
such that it satisfies the integral equation 
\begin{equation}\label{weakgCHN}
\begin{aligned}
0=\iint_{-\infty}^{+\infty} & \Big(
(\psi-\psi_{xx})u_t + (a\psi-c\psi_{xx})u^p u_x 
+ \tfrac{1}{2}(b-3pc)\psi_x u^{p-1} u_x^2 
\\&\qquad
+ \tfrac{1}{2}(p-1)(b-pc)\psi u^{p-2} u_x^3 
\Big) dx\; dt
\end{aligned}
\end{equation}
for all test functions $\psi(t,x)$ in $C^\infty_0((-T,T)\times\Rnum)$. 

\subsection{Multi-peakon solution}

Both the Camassa-Holm and Novikov equations possess multi-peakon solutions 
\cite{CamHol,HonWan,HonLunSzm}
which are a linear superposition of peaked travelling waves with time-dependent amplitudes and positions.
The form of these solutions is given by 
\begin{equation}\label{Npeakon}
u(t,x) = \sum_{i=1}^N \alpha_i(t)\exp(-|x-\beta_i(t)|),
\quad
N=1,2,\ldots
\end{equation}
where the amplitudes $\alpha_i(t)$ and positions $\beta_i(t)$ satisfy 
a Hamiltonian system of ODEs 
\begin{equation}\label{Npeakoneom}
\alpha_i{}'=\{\alpha_i,H\},
\quad
\beta_i{}'=\{\beta_i,H\},
\quad
i=1,\ldots,N
\end{equation}
given in terms of the Hamiltonian function
\begin{equation}\label{NpeakonH}
H = \tfrac{1}{2}\sum_{j,k=1}^N \alpha_j\alpha_k\exp(-|\beta_i-\beta_j|) . 
\end{equation}
The Poisson bracket $\{f,g\}$ in this system \eqref{Npeakoneom}
arises from the respective Hamiltonian operator formulations 
\cite{CamHol,HonWan} of these two equations
and has the standard canonical form in the case of Camassa-Holm equation
and a certain non-canonical form in the case of the Novikov equation.

We now investigate whether equation \eqref{4parmCHN} 
also admits multi-peakon solutions. 
It will be convenient to use the notation 
\begin{equation}\label{Npeakonu}
u = \sum_{i} \alpha_i e^{-|z_i|},
\quad
z_i=x-\beta_i
\end{equation}
where the summation is understood to go from $1$ to $N$. 
Note that the $x$-derivatives of $u$ are given by 
\begin{equation}\label{Npeakonux}
u_x = -\sum_{i} \sgn(z_i)\alpha_i e^{-|z_i|}
\end{equation}
and 
\begin{equation}\label{Npeakonuxx}
u_{xx} = \sum_{i} (-2\delta(z_i) +\sgn(z_i)^2)\alpha_i e^{-|z_i|}
\end{equation}
in terms of the sign function 
\begin{equation}
\sgn(z)=\begin{cases}
1 & z>0\\
-1 & z<0\\
0 & z=0
\end{cases}
\end{equation}
and the Dirac delta distribution
\begin{equation}\label{delta}
\delta(z)=\frac{d\,(\tfrac{1}{2}\sgn(z))}{dz}
\end{equation}
which has the properties 
$\delta(z)=0$ for $z\neq 0$,
and $\int_{-\epsilon}^{\epsilon}\delta(z)dz = 1$ for all $\epsilon >0$. 

To begin, 
we substitute the general multi-peakon expression \eqref{Npeakonu}
into the integral equation \eqref{weakgCHN}. 
There are two ways we can then proceed. 
One way is to assume $\beta_1<\beta_2<\cdots<\beta_N$ at a fixed $t>0$, 
split up the integral over $x$ into corresponding intervals,
and integrate by parts, 
similarly to the derivation of the single peakon solution. 
Another way, which is simpler, is to employ the following result from 
distribution theory \cite{GelShi}.

Let $f(x)$ be a piecewise $C^1$ function having at most jump discontinuities
at a finite number of points $x=x_i$ in $\Rnum$. 
Then, for any test function $\psi(x)$, 
\begin{equation}\label{ibp}
\int_{-\infty}^{\infty} \psi' f\;dx 
= -\sum_i \psi(x_i)[f]_{x_i} - \int_{-\infty}^{\infty} \psi \ns{f'}\;dx 
\end{equation}
where 
\begin{equation}\label{jumpf}
[f]_{x_i} = f(x_i{}^+) - f(x_i{}^-)
\end{equation}
is the jump in $f(x)$ at the point $x=x_i$,
and 
\begin{equation}\label{suppf}
\ns{f'} = 
\begin{cases} 
f' & x\neq x_i\\
0 & x=x_i
\end{cases}
\end{equation}
is the non-singular part of the distributional derivative of $f(x)$. 
We will now use this integration by parts relation \eqref{ibp}
to evaluate each term in the integral equation \eqref{weakgCHN}. 

The first term in equation \eqref{weakgCHN} yields
\begin{equation}
\begin{aligned}
\iint_{-\infty}^{+\infty} (\psi u_t-\psi_{xx} u_t)\; dx\; dt 
& = \iint_{-\infty}^{+\infty} \psi( u_t-\ns{u_{txx}})\; dx\; dt 
\\&\qquad
+ \sum_{i}\int_{-\infty}^{+\infty} (\psi_x(t,\beta_i)[u_t]_{\beta_i} - \psi(t,\beta_i)[u_{tx}]_{\beta_i} )\; dt . 
\end{aligned}
\end{equation}
From expression \eqref{Npeakonuxx}, 
we see 
\begin{equation}\label{nonsinguxx}
\ns{u_{xx}} = \sum_{i} \sgn(z_i)^2\alpha_i e^{-|z_i|} 
= u \quad\text{for}\quad x\neq \beta_i , 
\end{equation}
so thus $u_t-\ns{u_{txx}}= 0$ holds a.\ e.\ in $(-\infty,\infty)$. 
Hence 
\begin{equation}
\iint_{-\infty}^{+\infty} \psi( u_t-\ns{u_{txx}})\; dx\; dt =0
\end{equation}
and thus we get 
\begin{equation}\label{1stterm}
\iint_{-\infty}^{+\infty} (\psi u_t-\psi_{xx} u_t)\; dx\; dt 
= \sum_{i}\int_{-\infty}^{+\infty} ( \psi_x(t,\beta_i)[u_t]_{\beta_i} - \psi(t,\beta_i)[u_{tx}]_{\beta_i} )\; dt . 
\end{equation}

Next,
the second term in equation \eqref{weakgCHN} gives
\begin{equation}
\begin{aligned}
\iint_{-\infty}^{+\infty} (a\psi u^p u_x-c\psi_{xx} u^p u_x)\; dx\; dt 
& = \iint_{-\infty}^{+\infty} ( a\psi u^p u_x +c\psi_x (u^p \ns{u_{xx}} +pu^{p-1}\ns{u_x^2}) )\; dx\; dt 
\\&\qquad
+ \sum_{i}\int_{-\infty}^{+\infty} \psi_x(t,\beta_i)[u^p u_x]_{\beta_i} \; dt . 
\end{aligned}
\end{equation}
We now simplify the two parts of the integral involving $\psi_x$. 
For the first part, we have
\begin{equation}\label{2ndterm-integral1}
\iint_{-\infty}^{+\infty} \psi_x u^p \ns{u_{xx}} \; dx\; dt 
= \iint_{-\infty}^{+\infty} \psi_x u^{p+1} \; dx\; dt 
= -\iint_{-\infty}^{+\infty} \psi (p+1)u^p u_x \; dx\; dt 
\end{equation}
after using relation \eqref{nonsinguxx} and then integrating by parts. 
For the second part, 
since $\ns{u_x^2}= u_x^2$ holds a.\ e.\ in $(-\infty,\infty)$, 
we have 
\begin{equation}\label{2ndterm-part2}
\begin{aligned}
\iint_{-\infty}^{+\infty} \psi_x u^{p-1}\ns{u_x^2} \; dx\; dt 
& = \iint_{-\infty}^{+\infty} \psi_x u^{p-1}u_x^2 \; dx\; dt 
\\
& = \sum_{i}\int_{-\infty}^{+\infty} \psi(t,\beta_i)[u^{p-1}u_x^2]_{\beta_i} \; dt 
- \iint_{-\infty}^{+\infty} \psi (u^{p-1}\ns{u_x^2})_x \; dx\; dt 
\end{aligned}
\end{equation}
from applying the integration by parts relation \eqref{ibp}. 
By simplifying $(u^{p-1}\ns{u_x^2})_x = {(p-1)}u^{p-2}\ns{u_x^3} + 2u^{p-1} \ns{u_xu_{xx}} = (p-1)u^{p-2}u_x^3 + 2u^p u_x$ a.\ e.\ 
with the use of relation \eqref{nonsinguxx}, 
we see 
\begin{equation}
\iint_{-\infty}^{+\infty} \psi (u^{p-1}\ns{u_x^2})_x \; dx\; dt 
= \iint_{-\infty}^{+\infty} \psi ( (p-1)u^{p-2}u_x^3 + 2u^p u_x )\; dx\; dt . 
\end{equation}
Hence the integral \eqref{2ndterm-part2} becomes
\begin{equation}\label{2ndterm-integral2}
\begin{aligned}
\iint_{-\infty}^{+\infty} \psi_x u^{p-1}\ns{u_x^2} \; dx\; dt 
& = - \iint_{-\infty}^{+\infty} \psi ( (p-1)u^{p-2}u_x^3 + 2u^p u_x )\; dx\; dt 
\\&\qquad
-\sum_{i}\int_{-\infty}^{+\infty} \psi(t,\beta_i)u(\beta_i)^{p-1} [u_x^2]_{\beta_i} \; dt . 
\end{aligned}
\end{equation}
Then we have
\begin{equation}\label{2ndterm}
\begin{aligned}
& \iint_{-\infty}^{+\infty} (a\psi u^p u_x-c\psi_{xx} u^p u_x)\; dx\; dt 
\\&
= \iint_{-\infty}^{+\infty} \psi ( (a-(3p+1)c)u^p u_x -cp(p-1)u^{p-2}u_x^3 )\; dx\; dt 
\\&\qquad
-\sum_{i}\int_{-\infty}^{+\infty} cp\psi(t,\beta_i)u(\beta_i)^{p-1}[u_x^2]_{\beta_i} \; dt 
+\sum_{i}\int_{-\infty}^{+\infty} c\psi_x(t,\beta_i)u(\beta_i)^p [u_x]_{\beta_i} \; dt . 
\end{aligned}
\end{equation}

Similarly, 
the third term in equation \eqref{weakgCHN} gives
\begin{equation}
\begin{aligned}
\iint_{-\infty}^{+\infty} \tfrac{1}{2}(b-3pc)\psi_x u^{p-1} u_x^2\; dx\; dt
& = \iint_{-\infty}^{+\infty} (3pc-b) \psi (u^{p-1}\ns{u_xu_{xx}} +\tfrac{1}{2}(p-1)u^{p-2}\ns{u_x^3}) \; dx\; dt 
\\&\qquad
+\sum_{i}\int_{-\infty}^{+\infty} \tfrac{1}{2}(3pc-b)\psi(t,\beta_i)[u^{p-1} u_x^2]_{\beta_i} \; dt . 
\end{aligned}
\end{equation}
We can simplify the integral involving $\psi$ by the same steps 
used for the previous integral. 
This yields
\begin{equation}
\iint_{-\infty}^{+\infty} \psi (u^{p-1}\ns{u_xu_{xx}} +\tfrac{1}{2}(p-1)u^{p-2}\ns{u_x^3}) \; dx\; dt 
= \iint_{-\infty}^{+\infty} \psi (u^p u_x +\tfrac{1}{2}(p-1)u^{p-2} u_x^3) \; dx\; dt .
\end{equation}
Hence we then have
\begin{equation}\label{3rdterm}
\begin{aligned}
\iint_{-\infty}^{+\infty} \tfrac{1}{2}(b-3pc)\psi_x u^{p-1} u_x^2\; dx\; dt
& = \iint_{-\infty}^{+\infty} (3pc-b) \psi (u^p u_x +\tfrac{1}{2}(p-1)u^{p-2} u_x^3) \; dx\; dt 
\\&\qquad
+\sum_{i}\int_{-\infty}^{+\infty} \tfrac{1}{2}(3pc-b)\psi(t,\beta_i)u(\beta_i)^{p-1}[u_x^2]_{\beta_i} \; dt .
\end{aligned}
\end{equation}

Finally, by combining the three terms \eqref{1stterm}, \eqref{2ndterm}, \eqref{3rdterm} 
with the fourth term in equation \eqref{weakgCHN}, 
we obtain 
\begin{equation}\label{allterms}
\begin{aligned}
0= & 
(a-b-c)\iint_{-\infty}^{+\infty} u^pu_x\psi dx\; dt
+\sum_{i} \int_{-\infty}^{+\infty} \psi_{x}(t,\beta_i) ( [u_t]_{\beta_i} + c u(t,\beta_i)^p[u_x]_{\beta_i} )\; dt
\\&\qquad 
+\sum_{i} \int_{-\infty}^{+\infty} \psi(t,\beta_i)( -[u_{tx}]_{\beta_i} +\tfrac{1}{2}(pc-b)u(t,\beta_i)^{p-1}[u_x^2]_{\beta_i} )\; dt . 
\end{aligned}
\end{equation}
The jump terms are evaluated by 
\begin{gather}
[u_t]_{\beta_i}  = 2\alpha_i\beta_i',
\quad
[u_x]_{\beta_i}  = -2\alpha_i
\\
[u_{tx}]_{\beta_i}  = \frac{d\,[u_x]_{\beta_i}}{dt} = -2\alpha_i'
\\
[u_x^2]_{\beta_i}  = 2u_x(\beta_i)[u_x]_{\beta_i} = -4\alpha_iu_x(\beta_i)
\end{gather}
which all follow directly from the expressions \eqref{Npeakonu} and \eqref{Npeakonux}. 
Thus we get 
\begin{equation}
\begin{aligned}
0= & 
(a-b-c)\iint_{-\infty}^{+\infty} u^pu_x\psi dx\; dt
+2\sum_{i} \int_{-\infty}^{+\infty} \psi_{x}(t,\beta_i)(\alpha_i\beta_i{}' -c\alpha_i u(t,\beta_i)^p )\; dt
\\&\qquad 
+2\sum_{i} \int_{-\infty}^{+\infty} \psi(t,\beta_i)(\alpha_i{}' + (b-pc)\alpha_i u(t,\beta_i)^{p-1} u_x(t,\beta_i) )\; dt . 
\end{aligned}
\end{equation}
This equation is satisfied for all test functions $\psi$ iff 
\begin{equation}
a=b+c,
\quad
p\geq 0,
\quad
\beta_i{}' =c u(t,\beta_i)^p,
\quad
\alpha_i{}' = (pc-b)\alpha_i u(t,\beta_i)^{p-1} u_x(t,\beta_i) 
\end{equation}
which determines the amplitudes $\alpha_i$ and the positions $\beta_i$ 
in the multi-peakon expression \eqref{Npeakonu}. 
Thus we have established the following result. 

\begin{prop}\label{class-Npeakon}
The integral equation \eqref{weakgCHN} admits an $N$-peakon solution \eqref{Npeakonu} for all $N\geq 1$ 
only in the case 
\begin{equation}
a=b+c,
\quad
p\geq0 . 
\end{equation}
\end{prop}

From Propositions~\ref{class-Npeakon} and~\ref{class-peakon}, 
we have a classification of all cases for which the 4-parameter equation \eqref{4parmCHN}
possesses both single peakon and multi-peakon solutions. 

\begin{thm}\label{class-peakons}
The 4-parameter equation \eqref{4parmCHN} admits 
single peakon and multi-peakon solutions iff 
\begin{equation}\label{Npeakon-case}
a=b+c,
\quad
c\neq0,
\quad
p\geq 0 . 
\end{equation}
In this case, a general $N$-peakon solution has the form \eqref{Npeakon},
where the amplitudes $\alpha_i(t)$ and positions $\beta_i(t)$ satisfy 
the system of ODEs 
\begin{align}
& 
\beta_i{}' =c \Big(\alpha_j +\sum_{\substack{j\neq i\\j=1}}^{N} \alpha_j\exp(-|\beta_{i,j}|)\Big)^p,
\label{pos-eom}\\
&
\alpha_i{}' = (b-pc)\alpha_i \Big(\alpha_j +\sum_{\substack{j\neq i\\j=1}}^{N} \alpha_j\exp(-|\beta_{i,j}|)\Big)^{p-1} \sum_{\substack{k\neq i\\k=1}}^{N} \sgn(\beta_{i,k})\alpha_k\exp(-|\beta_{i,k}|) ,
\label{ampl-eom}
\end{align}
in terms of the separations
\begin{equation}\label{sep}
\beta_{i,j} = \beta_i-\beta_j . 
\end{equation}
\end{thm}

This result generalizes related work in Ref.\cite{Him1} 
which established the existence of single and multi-peakon solutions 
for a 2-parameter equation defined by the case 
$a=b+1$, $c=1$ of equation \eqref{4parmCHNm}. 
(In particular, the derivation in Ref.\cite{Him1} was completely formal, 
whereas the steps here provide a rigorous proof applied to 
the more general 4-parameter equation \eqref{4parmCHNm}.)

It is easy to check that the general $N$-peakon ODE system \eqref{pos-eom}--\eqref{ampl-eom}
reduces to the well-known multi-peakon systems 
for the $b$-equation \eqref{CH-DP} when $(p,a,b,c)=(1,b+1,b,1)$,
which includes the the Camassa-Holm equation and the Degasperis-Procesi equation
when $b=2$ and $b=3$, respectively, 
as well as for the Novikov equation when $(p,a,b,c)=(2,4,3,1)$. 

\subsection{Constants of motion}

The ODE system \eqref{pos-eom}--\eqref{ampl-eom} for the amplitudes and positions of the $N$ peakons in the expression \eqref{Npeakon}
inherits constants of motion (i.e.\ time-independent quantities) 
given by the conserved integrals that are admitted by equation \eqref{4parmCHNm}
in the case \eqref{Npeakon-case}.  
From Theorem~\ref{class-loworderTX}, 
there are six conserved integrals \eqref{0thC-a}--\eqref{2ndC-2} 
which we can consider. 

The first conserved integral \eqref{0thC-a} yields
\begin{equation}\label{com-mass}
\begin{aligned}
& \mathcal C_1= \int_{-\infty}^{\infty} u\; dx 
= 2\sum_{i=1}^{N} \alpha_i =\const
\\
& \text{ iff }\quad
p=1
\quad\text{ or }\quad
b=pc . 
\end{aligned}
\end{equation}
This quantity $\mathcal C_1$ is the total mass for the $N$-peakon solution.
A weighted mass arises from the second conserved integral \eqref{0thC-c}, 
\begin{equation}\label{com-weightedmass}
\begin{aligned}
& \mathcal C_2= \int_{-\infty}^{\infty} e^{-\sqrt{a/c}\;|x|} u \; dx
= \frac{2c}{c-a} \sum_{i=1}^{N} (e^{-\sqrt{a/c}\;|\beta_i|} - e^{-|\beta_i|})\alpha_i 
=\const
\\
& \text{ iff }\quad
p=1, 
\quad 
a = 4c, 
\quad
b=3c .
\end{aligned}
\end{equation}
The next conserved integral \eqref{0thC-b} gives
\begin{equation}\label{com-H1norm}
\begin{aligned}
& \mathcal C_4= \int_{-\infty}^{\infty} u^2 + u_x^2 \; dx 
= \int_{-\infty}^{\infty} u(u- u_{xx}) \; dx 
= 2\sum_{i,j=1}^{N} \alpha_i\alpha_j e^{-|\beta_i-\beta_j|} 
=\const
\\
& \text{ iff }\quad
a=(p+2)c ,
\quad
b=(p+1)c 
\end{aligned}
\end{equation}
which is the $H^1$ norm of the $N$-peakon solution.

The fourth conserved integral \eqref{0thC-e} does not exist in the case \eqref{Npeakon-case},
since $a=b+c=c$ and $b=2c$ together imply that $a=b=c=0$. 
Last, the two conserved integrals \eqref{2ndC-1} and \eqref{2ndC-2}
are nonlinear in $u_{xx}$ which is a distribution. 
As a consequence, 
both these integrals are ill-defined for the the $N$-peakon solution.

\section{Unified family of Camassa-Holm-Novikov equations}
\label{unifiedeqn}

From Theorems~\ref{class-loworderTX} and~\ref{class-peakons},
the low-order conservation laws \eqref{T1-CH-N}--\eqref{T2-CH-N}
as well as the $N$-peakon solution expression \eqref{Npeakon}
of the Camassa-Holm and Novikov equations 
are admitted simultaneously by the 4-parameter equation \eqref{4parmCHNm}
iff its parameters $(\tilde a,b,c,p)$ satisfy 
\begin{equation}
\tilde a =0,
\quad
b=(p+1)c,
\quad
c\neq0,
\quad
p\geq 0 . 
\end{equation}
After a scaling transformation $t\rightarrow t/c$ is used to put $c=1$, 
equation \eqref{4parmCHNm} reduces to the 1-parameter gCHN equation \eqref{CHNm}
presented in section~\ref{intro}. 

\subsection{Dynamics of multi-peakon solutions}

The explicit system describing $N$-peakon solutions of the gCHN equation \eqref{CHNm} for all $p\geq 0$ 
is given by 
\begin{align}
& 
\beta_i{}' =\Big(\alpha_j +\sum_{\substack{j\neq i\\j=1}}^{N} \alpha_j\exp(-|\beta_{i,j}|)\Big)^p ,
\quad
\beta_{i,j} = \beta_i -\beta_j,
\label{peakon-pos-eom}\\
&
\alpha_i{}' = \alpha_i \Big(\alpha_j +\sum_{\substack{j\neq i\\j=1}}^{N} \alpha_j\exp(-|\beta_{i,j}|)\Big)^{p-1} \sum_{\substack{k\neq i\\k=1}}^{N} \sgn(\beta_{i,k})\alpha_k\exp(-|\beta_{i,k}|)  ,
\label{peakon-ampl-eom}
\end{align}
where $\alpha_i(t)$ and $\beta_i(t)$ are, respectively, 
the amplitudes and positions appearing in the general $N$-peakon expression \eqref{Npeakon}. 
The $H^1$ norm \eqref{com-H1norm} of the $N$-peakon solution provides 
a constant of motion 
\begin{equation}\label{peakon-H1norm}
H = \sum_{i,j=1}^{N} \alpha_i\alpha_j e^{-|\beta_{i,j}|} 
=\const \geq 0
\end{equation}
which is determined by the initial amplitudes and initial separations. 

When all of the amplitudes are positive, $\alpha_i>0$, for all $t\geq 0$, 
the solution expression \eqref{Npeakon} is a superposition of $N\geq1$ peakons,
each of which is right moving. 
In this case, 
the constant of motion \eqref{peakon-H1norm} directly gives the inequality
$H>\alpha_i>0$, $i=1,2,\ldots,N$, 
which implies that any collisions among the $N$ peakons are elastic. 

When all of the amplitudes are negative, $\alpha_i<0$, for all $t\geq 0$, 
the solution expression \eqref{Npeakon} is instead a superposition of $N\geq1$ anti-peakons,
each of which is either right moving if $p$ is even or left moving if $p$ is odd. 
Similarly to the previous case, 
the constant of motion \eqref{peakon-H1norm} yields 
$H>|\alpha_i|>0$, $i=1,2,\ldots,N$, 
implying that any collisions among the $N$ anti-peakons are elastic. 

In the case when some amplitudes have opposite signs, 
or an amplitude changes its sign at some $t>0$,
the solution expression \eqref{Npeakon} then describes a superposition of both peakons and anti-peakons. 
Although the constant of motion is still non-negative, 
the amplitudes are no longer bounded by $H\geq0$. 
As a consequence, wave breaking can occur in collisions,
which we will now show for the case $N=2$. 

\subsection{Wave breaking in collisions between peakons and anti-peakons}

For $N=2$, the system \eqref{peakon-pos-eom}--\eqref{peakon-ampl-eom}
describing $2$-peakon solutions 
\begin{equation}\label{2peakonu}
u = \alpha_1 e^{-|x-\beta_1|} + \alpha_2 e^{-|x-\beta_2|} 
\end{equation}
takes a simple form. 
First, the constant of motion \eqref{peakon-H1norm} can be used to 
express the relative separation $|\beta_{1,2}|=|\beta_1-\beta_2|$ in terms of 
the two amplitudes $\alpha_1$ and $\alpha_2$ through the relation 
\begin{equation}\label{2peakon-H}
e^{-|\beta_{1,2}|} = \frac{H -\alpha_1^2 -\alpha_2^2}{2\alpha_1\alpha_2} . 
\end{equation}
Then, the equations of motion for the two positions $\beta_1$ and $\beta_2$ 
and the two amplitudes $\alpha_1$ and $\alpha_2$ are given by 
\begin{gather}
\beta_1{}' = A_1^p, 
\label{2peakon-pos1-eom}\\
\beta_2{}' = A_2^p, 
\label{2peakon-pos2-eom}\\
\alpha_1{}' = \tfrac{1}{2} \sgn(\beta_{1,2}) A_1^{p-1}(H -\alpha_1^2 -\alpha_2^2), 
\label{2peakon-ampl1-eom}\\
\alpha_2{}' = -\tfrac{1}{2} \sgn(\beta_{1,2}) A_2^{p-1}(H -\alpha_1^2 -\alpha_2^2),
\label{2peakon-ampl2-eom}
\end{gather}
with
\begin{equation}\label{A1A2}
A_1 = \frac{H +\alpha_1^2 -\alpha_2^2}{2\alpha_1},
\quad
A_2 = \frac{H +\alpha_2^2 -\alpha_1^2}{2\alpha_2},
\end{equation}
and
\begin{equation}
\beta_{1,2} = \beta_1-\beta_2 . 
\end{equation}
If another constant of motion could be found for this system,
then the system could be reduced to two separated ODEs for the two amplitudes,
plus two quadratures for the two positions,
which would allow the general solution to be obtained. 
Even without another constant of motion, 
it is still possible to do a qualitative analysis of all solutions by 
studying the phase plane $(\alpha_1,\alpha_2)$ of the coupled ODEs \eqref{2peakon-ampl1-eom}--\eqref{2peakon-ampl2-eom}
for the amplitudes. 

We start from the relation \eqref{2peakon-H}, 
which imposes inequalities on the amplitudes, 
\begin{equation}\label{H-rels}
0\leq \frac{H -\alpha_1^2 -\alpha_2^2}{2\alpha_1\alpha_2} \leq 1 . 
\end{equation}
For a given value of $H>0$, these two inequalities define the domain for
all $2$-peakon solutions in the phase plane $(\alpha_1,\alpha_2)$.
The boundary of the domain corresponds to the two equalities
\begin{equation}\label{H-circ}
\alpha_1^2 +\alpha_2^2 = H, 
\quad
|\beta_{1,2}| =\infty
\end{equation}
and
\begin{equation}\label{H-lines}
(\alpha_1 +\alpha_2)^2 =H,
\quad
|\beta_{1,2}| =0
\end{equation}
which consist of a circle and two parallel lines. 
The circle comprises the equilibrium points of the amplitude ODEs \eqref{2peakon-ampl1-eom}--\eqref{2peakon-ampl2-eom} in the phase plane. 
Each point on the circle is a limit of a $2$-peakon solution 
describing an asymptotic superposition of two $1$-peakon solutions, 
in which the amplitudes are constant and the positions are infinitely separated.
The lines each constitute a degenerate $2$-peakon solution 
in which the two positions coincide and the sum of the two amplitudes is constant,
describing a peakon solution 
\begin{equation}
u(t,x) = \sqrt{H}\exp(-|x-\sqrt{H}^p t|)
\end{equation}
in the case of the upper line, 
and an anti-peakon solution 
\begin{equation}
u(t,x)= -\sqrt{H}\exp(-|x-(-\sqrt{H})^p t|)
\end{equation}
in the case of the lower line. 

The entire solution domain divides into four parts 
which are related by a reflection symmetry 
$(\alpha_1,\alpha_2) \longleftrightarrow (-\alpha_1,-\alpha_2)$. 
One part of the domain is given by the points lying between the circle \eqref{H-circ} and the upper line \eqref{H-lines} in the first quadrant, 
which comprises all solutions describing two peakons. 
There is a counterpart given by the points lying between the circle \eqref{H-circ} and the lower line \eqref{H-lines} in the third quadrant, 
which comprises all solutions describing two anti-peakons. 
The two other parts of the domain comprise all solutions describing 
a peakon and an anti-peakon. 
These parts are given by the points between the segments of the upper and lower lines that lie outside of the circle. 

Within this solution domain in the phase plane, 
the flow defined by the amplitude ODEs \eqref{2peakon-ampl1-eom}--\eqref{2peakon-ampl2-eom} 
depends on the nonlinearity power $p$ and the sign of the separation $\beta_{1,2}$. 
We are interested in flows that describe a collision between a peakon and an anti-peakon. 
This condition can be used to determine $\sgn(\beta_{1,2})$ at $t=0$ 
at each point in the phase plane by considering the ODE
\begin{equation}\label{sep-eom}
\beta_{1,2}{}' = A_1^p - A_2^p
\end{equation}
for the separation. 
If $\beta_{1,2}{}' >0$, then the relative separation $|\beta_{1,2}|$ 
between the peakon and anti-peakon will be decreasing only if $\beta_{1,2}<0$. 
Similarly, if $\beta_{1,2}{}' <0$, then the relative separation $|\beta_{1,2}|$ 
between the peakon and anti-peakon will be decreasing only if $\beta_{1,2}>0$.  
Hence, a necessary condition for a collision to occur is that 
$\beta_{1,2}{}'$ and $\beta_{1,2}$ have opposite signs during the flow. 
Since $\beta_{1,2}=0$ can occur only on the upper and lower lines \eqref{H-lines},
which are boundaries of the domain in which solutions describe a collision between a peakon and an anti-peakon, 
we can impose 
\begin{equation}\label{sep-cond}
\sgn(\beta_{1,2}) = \sgn(A_2^p - A_1^p) 
\end{equation}
at each point in the phase plane. 
Note $\sgn(\beta_{1,2})=0$ holds iff $A_1=A_2$ when $p$ is odd, 
and $A_1=\pm A_2$ when $p$ is even. 
The points given by $A_1=A_2$ in the phase plane consist of the lines \eqref{H-lines} and $\alpha_1=\alpha_2$, 
while the points given by $A_1=-A_2$ consist of the lines that are 
perpendicular to each of those three lines. 
Consequently, hereafter we will consider initial conditions  
\begin{equation}\label{ampl-sign}
\alpha_2(0)>0 \quad\text{(peakon)}
\quad\text{and}\quad
\alpha_1(0)<0 \quad\text{(anti-peakon)}
\end{equation}
and 
\begin{equation}\label{ampl-cond}
\alpha_2(0) + \alpha_1(0)>0 
\end{equation}
without loss of generality. 
(Note that reversing the sign in the initial condition \eqref{ampl-cond}
will correspond to reflecting the flow about the line $\alpha_1+ \alpha_2=0$
in the phase plane.)

Under the collision condition \eqref{sep-cond} and initial conditions \eqref{ampl-sign}--\eqref{ampl-cond}, 
the flow then depends only on the nonlinearity power $p$. 
The case $p=1$, which represents the Camassa-Holm equation, is special, 
since there is another constant of motion $M = \alpha_1+\alpha_2=\const$
which is given by the total mass \eqref{com-mass}. 
This implies that the flow simply consists of parallel lines
in the phase plane. 
In all other cases $p\neq 1$, the flow is no longer given by straight lines 
and has a much richer structure. 

\begin{figure}
\centering
\includegraphics[width=.9\textwidth]{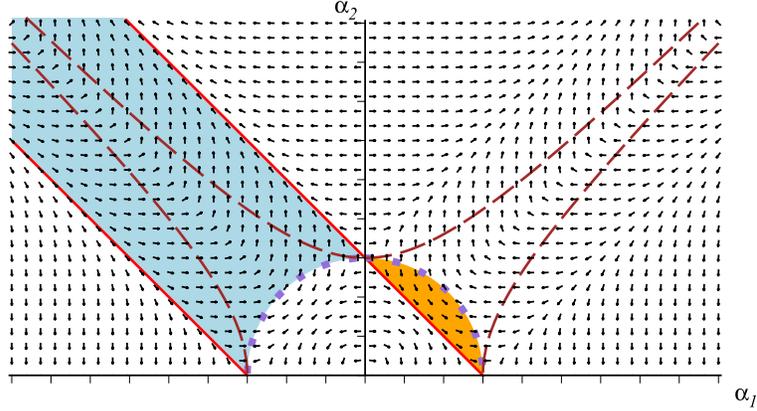}
\caption{Phase plane for collision of peakon and anti-peakon when $p=4$}
\label{p=4-plane}
\end{figure}

\begin{figure}
\centering
\includegraphics[width=.9\textwidth]{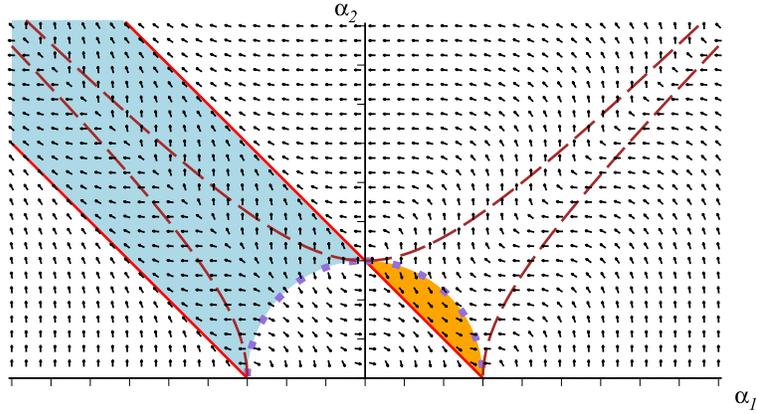}
\caption{Phase plane for collision of peakon and anti-peakon when $p=3$}
\label{p=3-plane}
\end{figure}

The flows for all even powers $p=2,4,\ldots$ are qualitatively similar to 
the case $p=2$, which represents the Novikov equation. 
A picture of the phase plane for $p=4$ is shown in Fig.~\ref{p=4-plane}.
Clearly, in the second quadrant, 
the upper line $\alpha_1+\alpha_2=\sqrt{H}$ is a stable asymptotic attractor 
for solutions describing a peakon ($\alpha_2>0$) and an anti-peakon ($\alpha_1<0$),
while the lower line $\alpha_1+\alpha_2=-\sqrt{H}$ is an unstable asymptotic attractor. 
In the fourth quadrant, these behaviours are reversed. 

The flows for all other odd powers $p=3,5,\ldots$ are qualitatively similar to 
the case $p=3$ which is shown in Fig.~\ref{p=3-plane}.
In the second quadrant, 
both the upper and lower lines $\alpha_1+\alpha_2=\pm\sqrt{H}$ are stable asymptotic attractors 
for solutions describing a peakon ($\alpha_2>0$) and an anti-peakon ($\alpha_1<0$). 
The line $\alpha_1+\alpha_2=0$ is an unstable asymptotic attractor. 
In the fourth quadrant, the behaviour is the same. 

\begin{figure}[t]
\centering
\begin{subfigure}[t]{.35\textwidth}
\includegraphics[width=\textwidth]{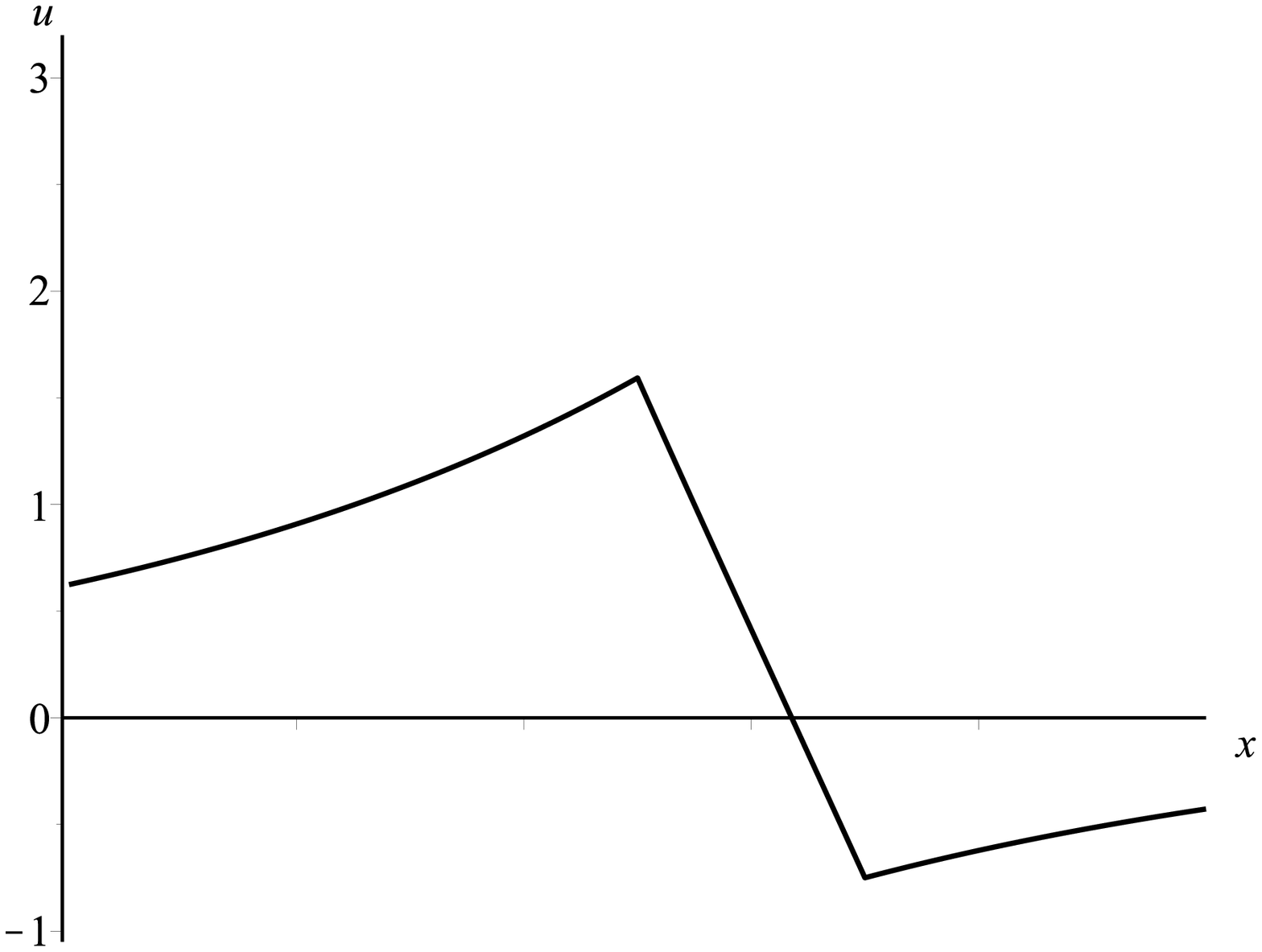}
\captionof{figure}{$t=0$}
\end{subfigure}%
\begin{subfigure}[t]{.35\textwidth}
\includegraphics[width=\textwidth]{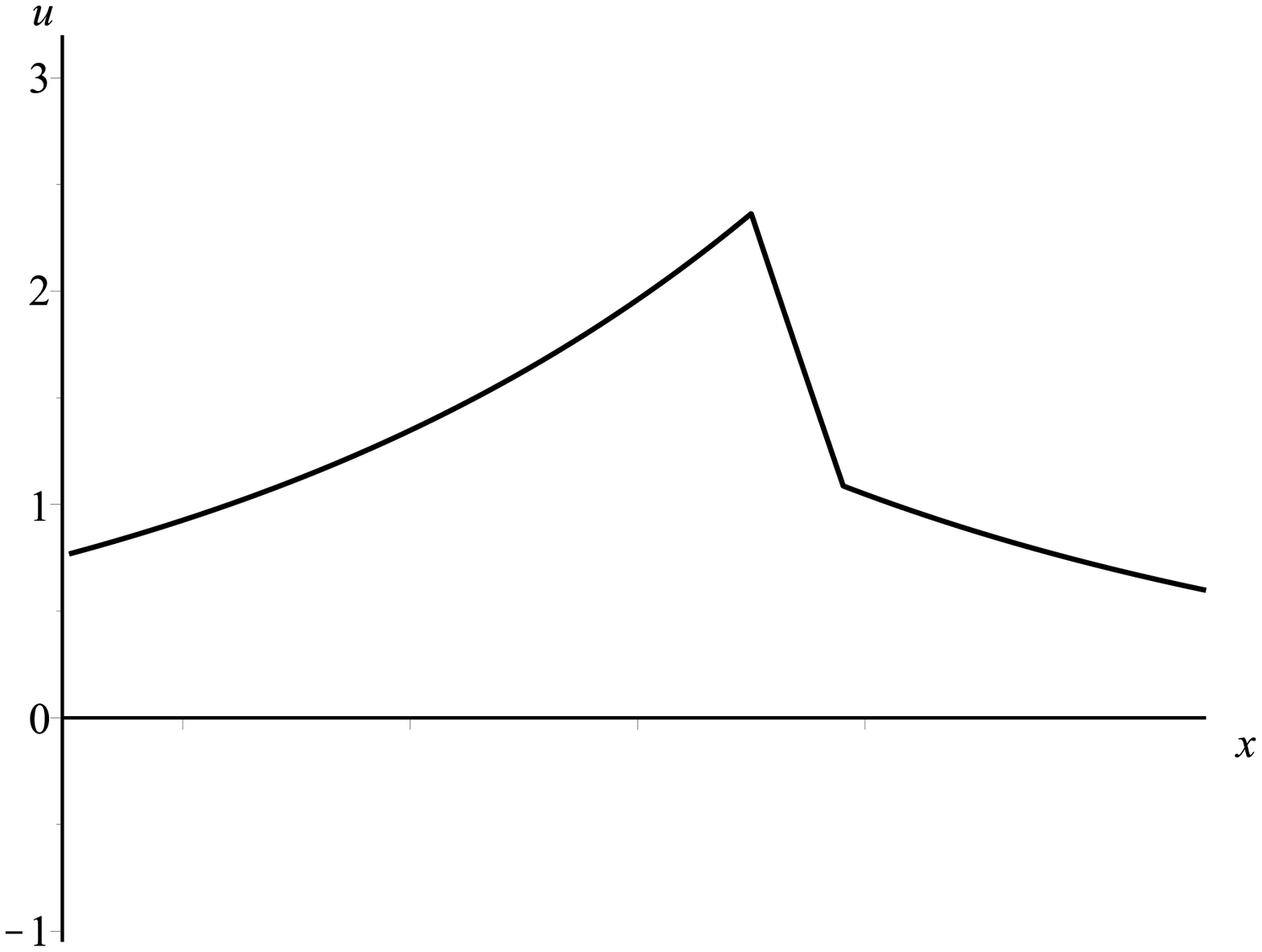}
\captionof{figure}{$t=0.5T$}
\end{subfigure}
\begin{subfigure}[t]{.35\textwidth}
\includegraphics[width=\textwidth]{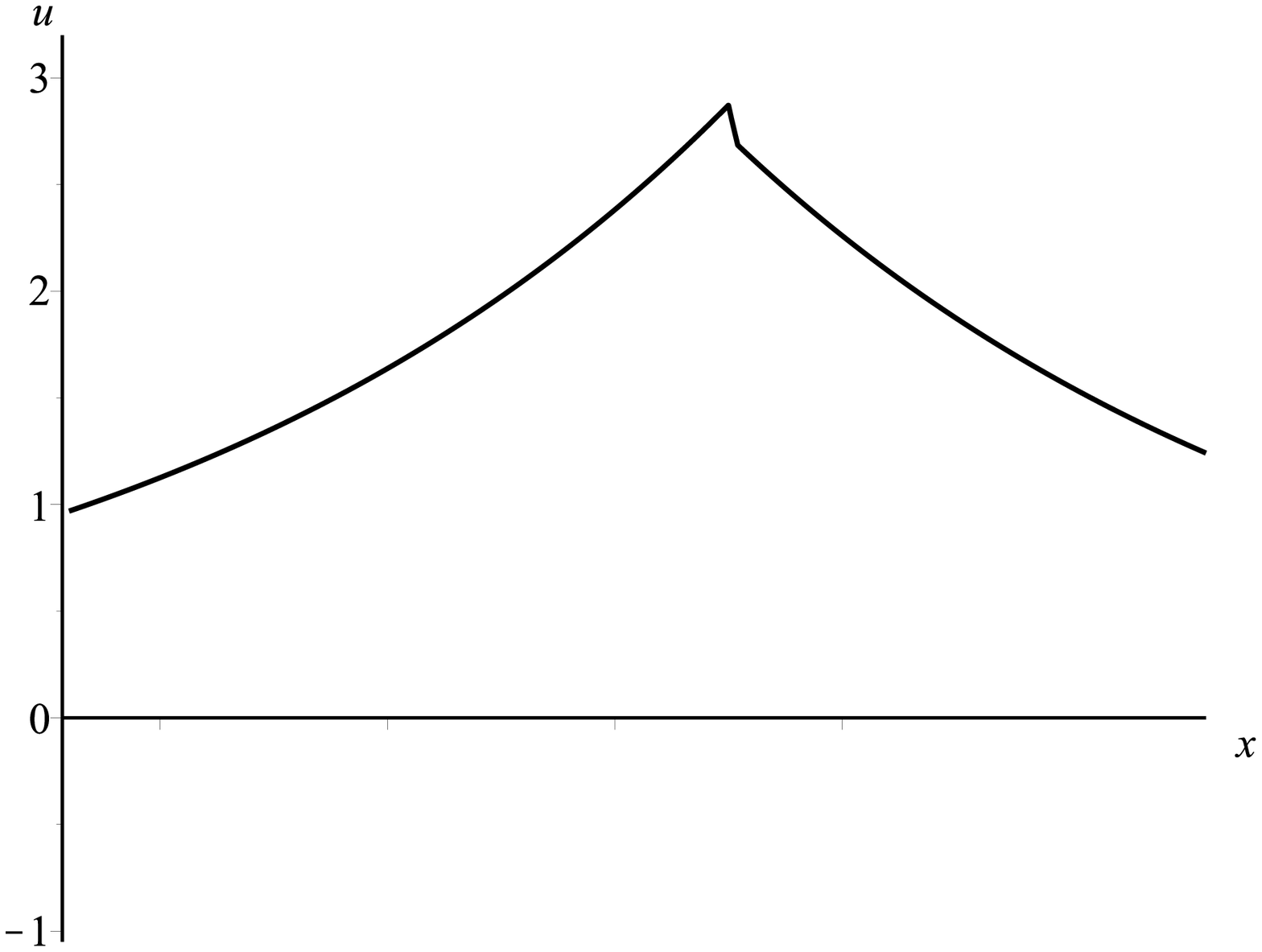}
\captionof{figure}{$t=0.72T$}
\end{subfigure}%
\begin{subfigure}[t]{.35\textwidth}
\includegraphics[width=\textwidth]{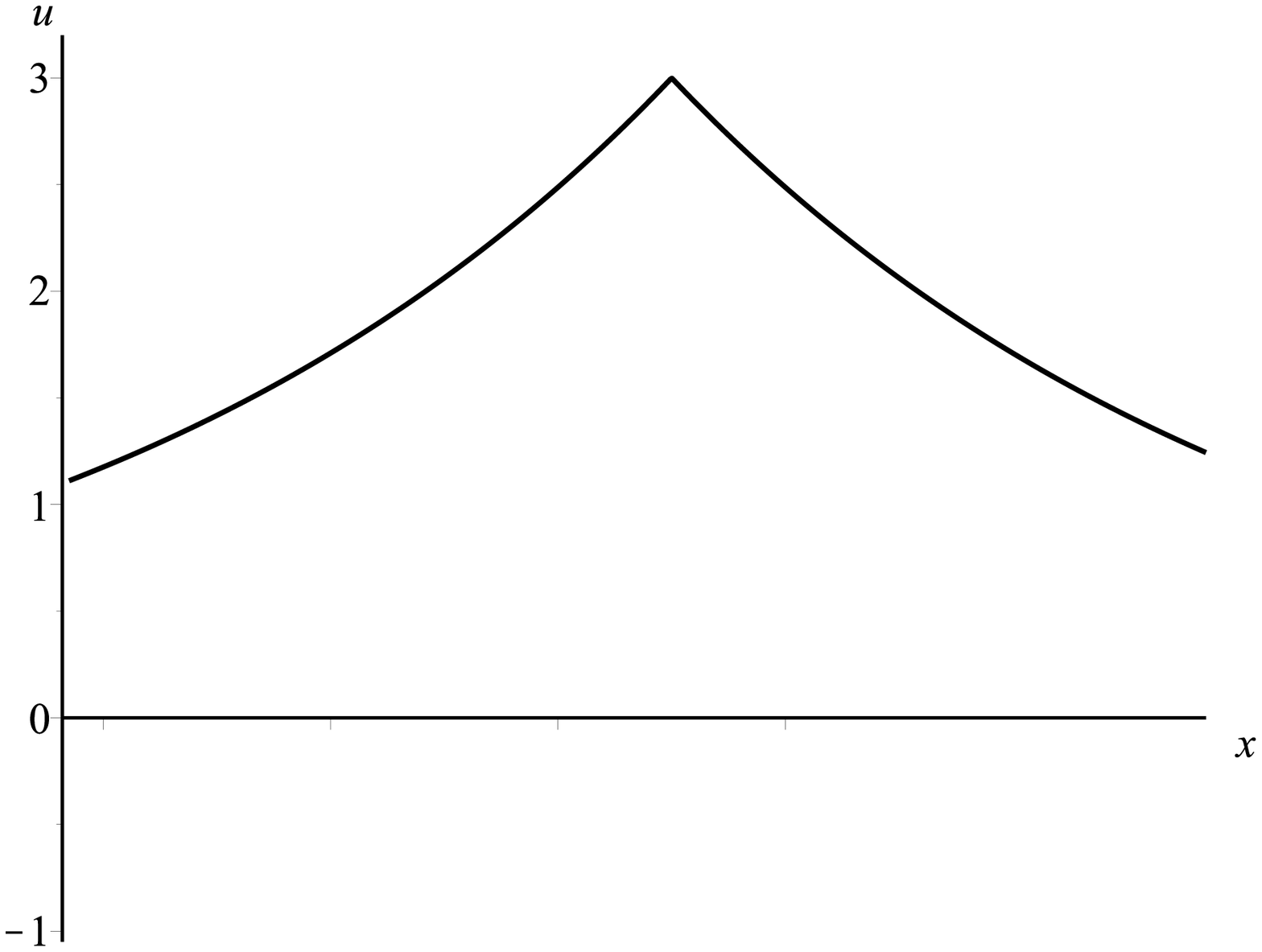}
\captionof{figure}{$t=T$}
\end{subfigure}
\caption{Collision of peakon and anti-peakon in the case $p=3$ ($\alpha_1(0)=-3.5$, $\alpha_2(0)=4$)}
\label{p=3-collide}
\end{figure}

\begin{figure}[t]
\centering
\begin{subfigure}[t]{.33\textwidth}
\includegraphics[width=\textwidth]{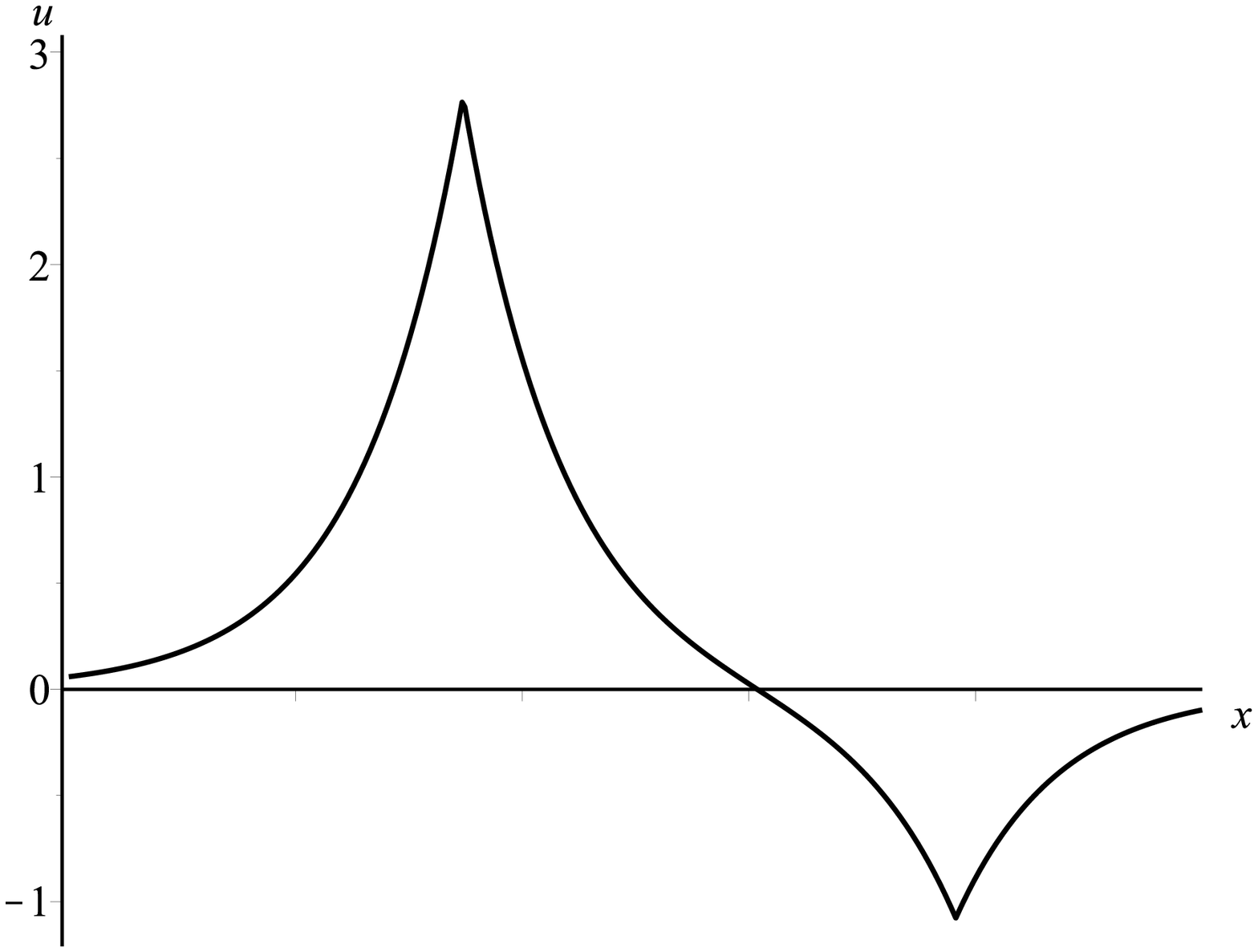}
\captionof{figure}{$t=0$}
\end{subfigure}%
\begin{subfigure}[t]{.33\textwidth}
\includegraphics[width=\textwidth]{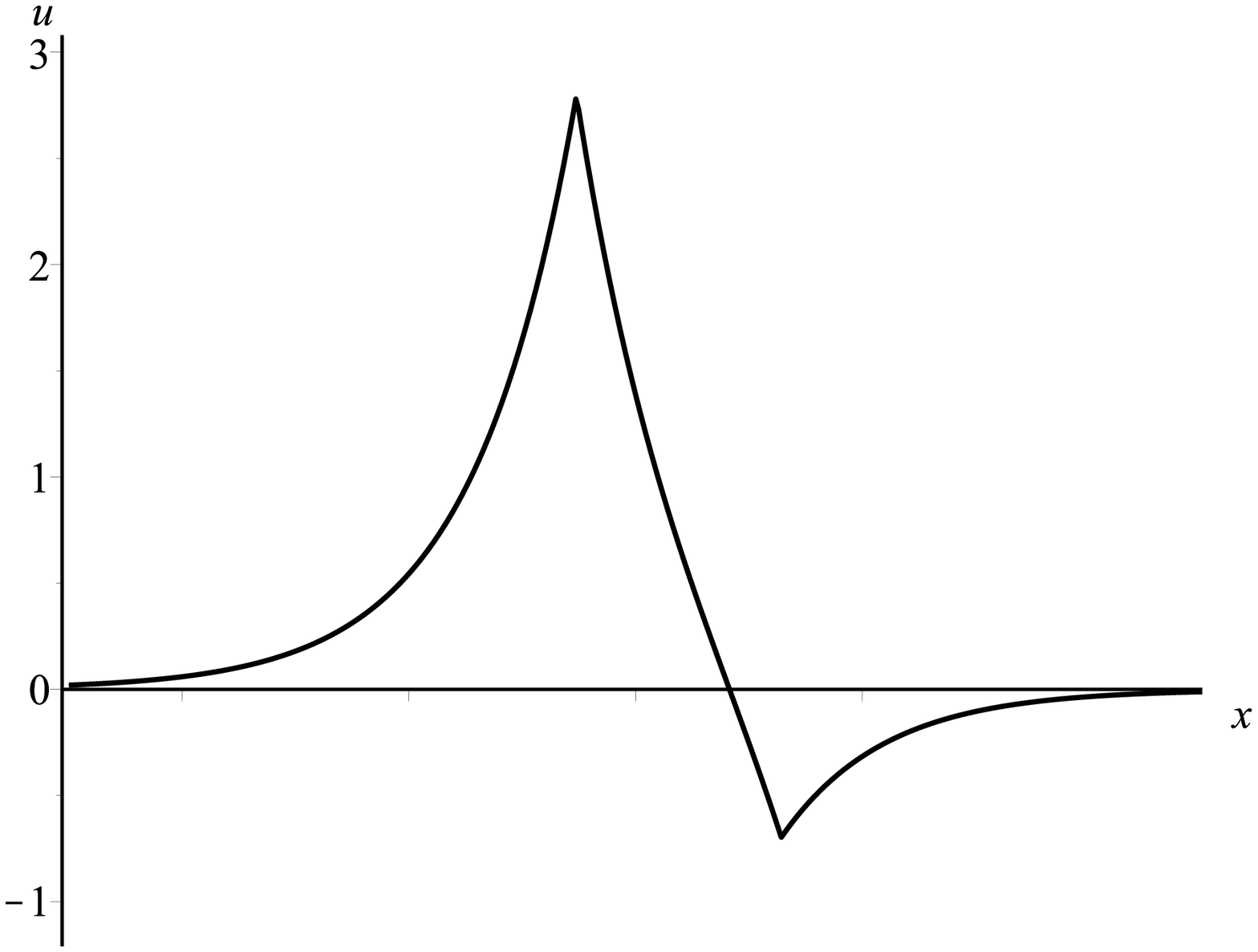}
\captionof{figure}{$t=0.5T$}
\end{subfigure}
\begin{subfigure}[t]{.33\textwidth}
\includegraphics[width=\textwidth]{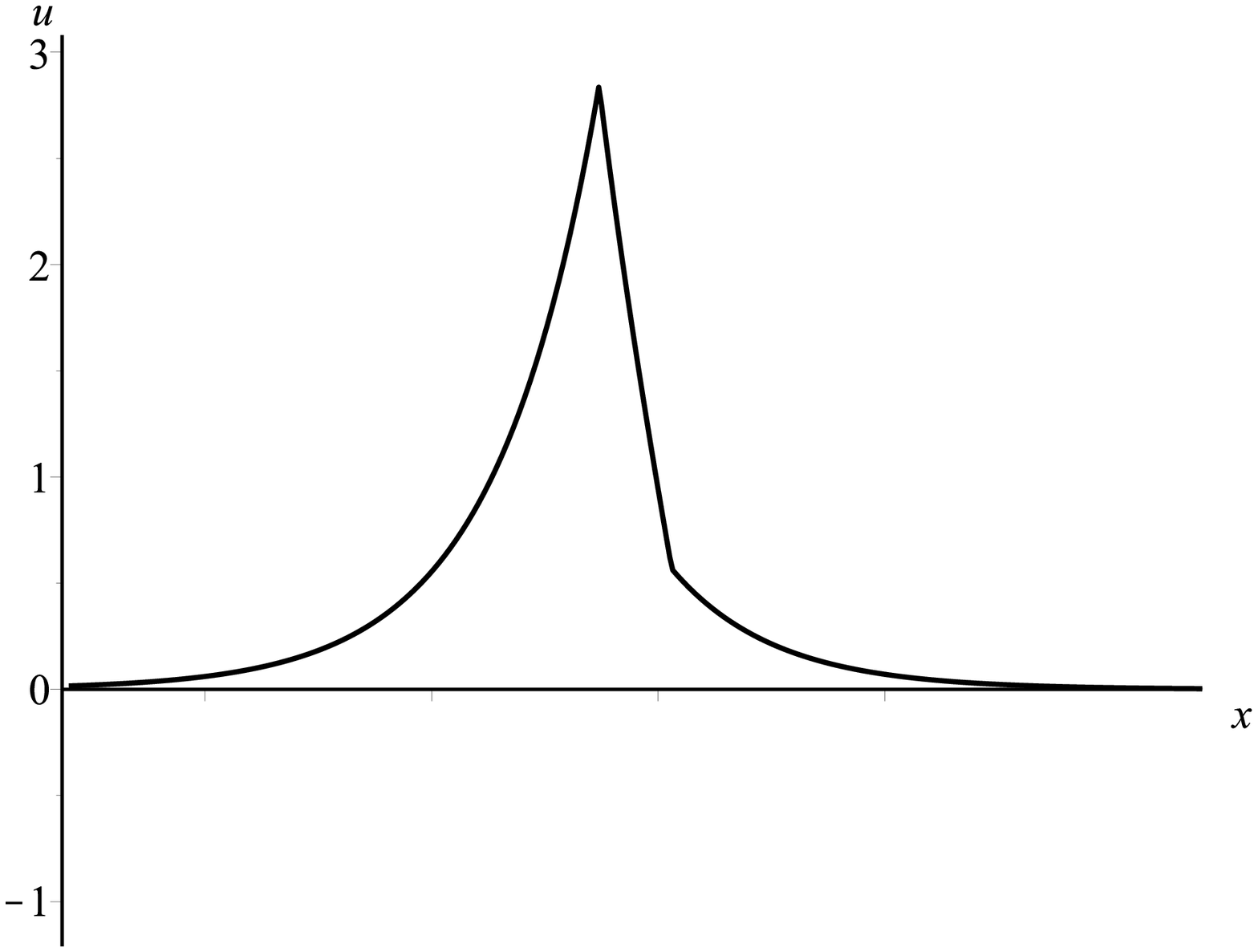}
\captionof{figure}{$t=0.72T$}
\end{subfigure}%
\begin{subfigure}[t]{.33\textwidth}
\includegraphics[width=\textwidth]{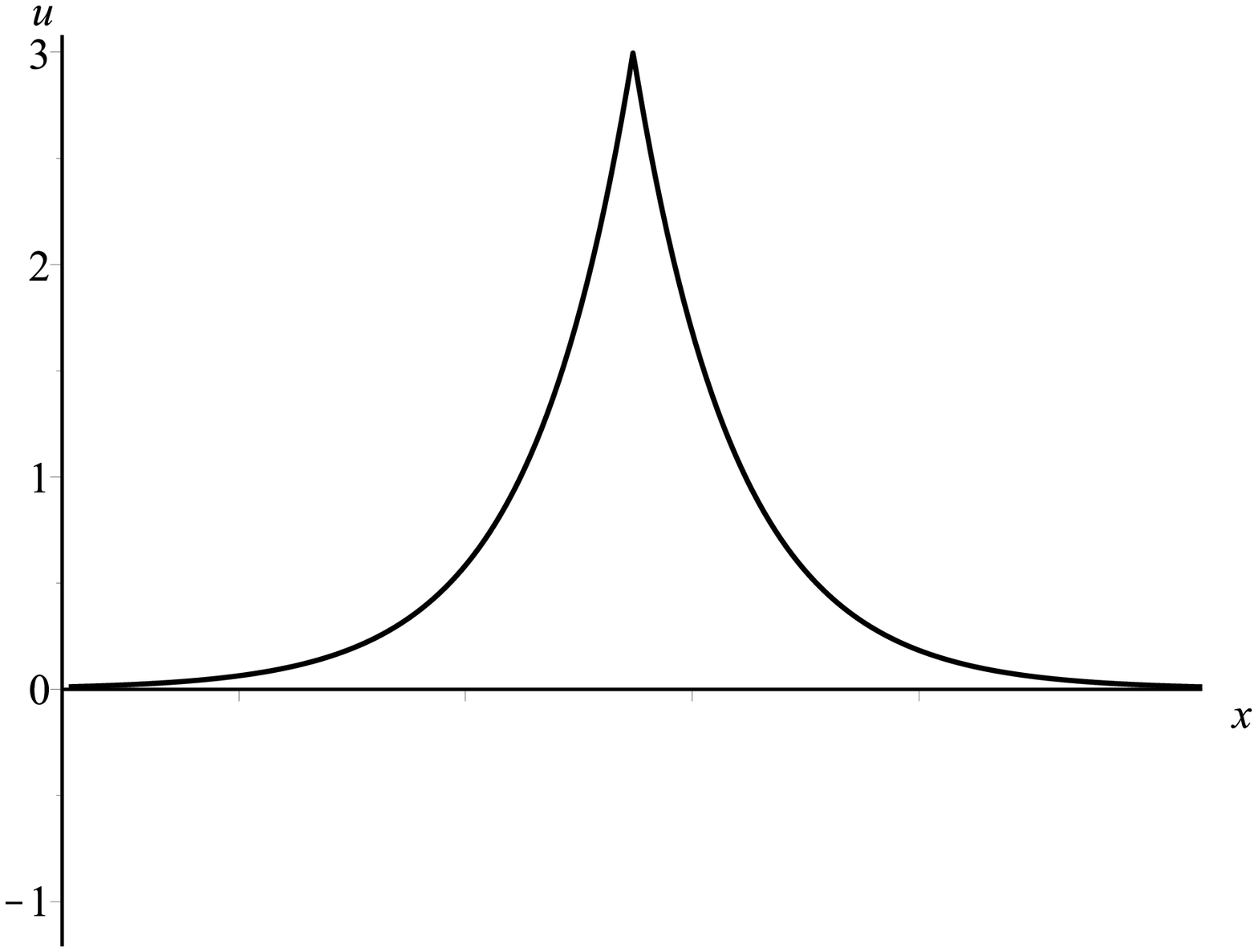}
\captionof{figure}{$t=T$}
\end{subfigure}
\caption{Collision of peakon and anti-peakon in the case $p=4$ ($\alpha_1(0)=-1.1$, $\alpha_2(0)=2.8$)}
\label{p=4-collide}
\end{figure}

\begin{figure}[H]
\centering
\begin{subfigure}[t]{.3\textwidth}
\includegraphics[width=\textwidth]{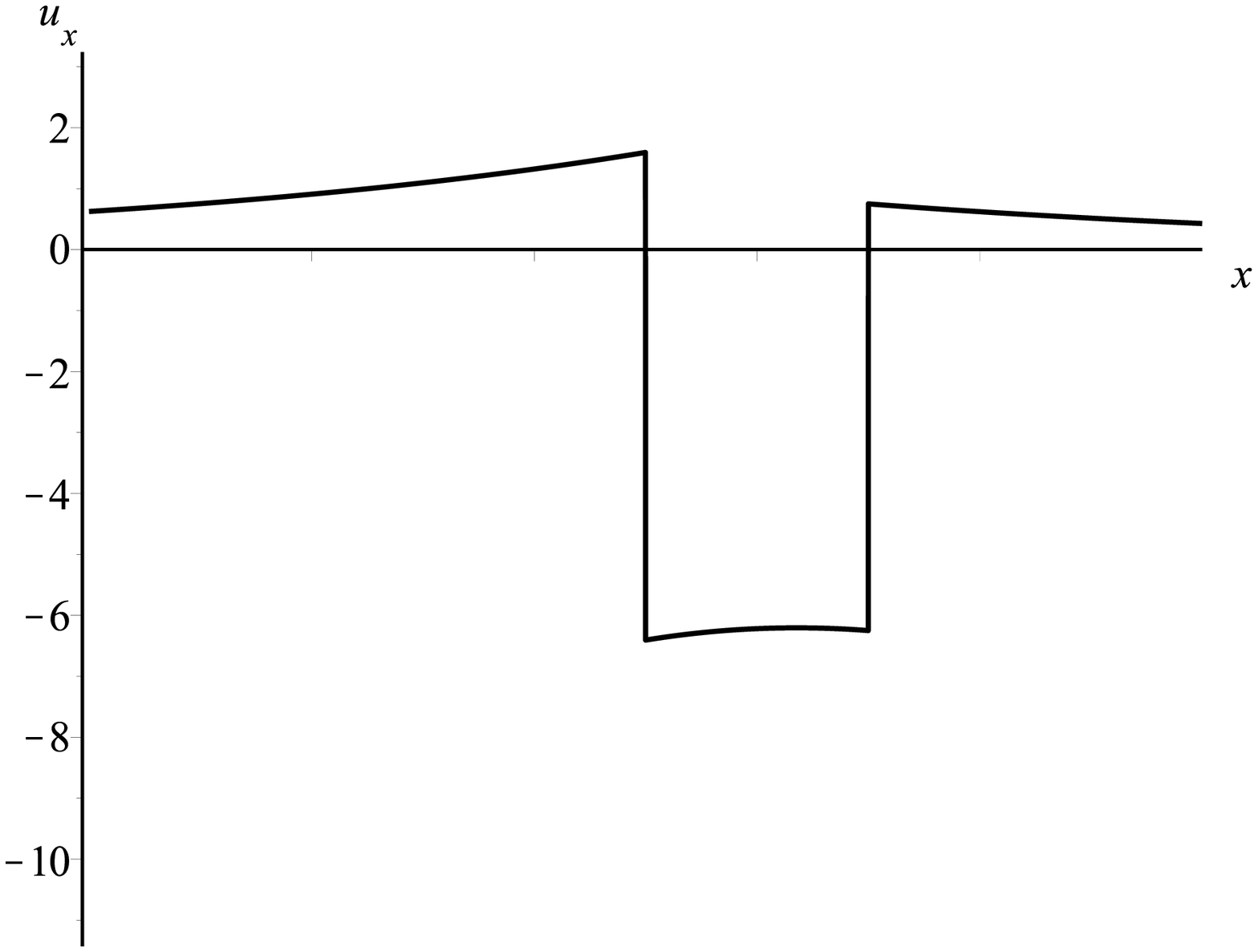}
\captionof{figure}{$t=0$}
\end{subfigure}%
\begin{subfigure}[t]{.3\textwidth}
\includegraphics[width=\textwidth]{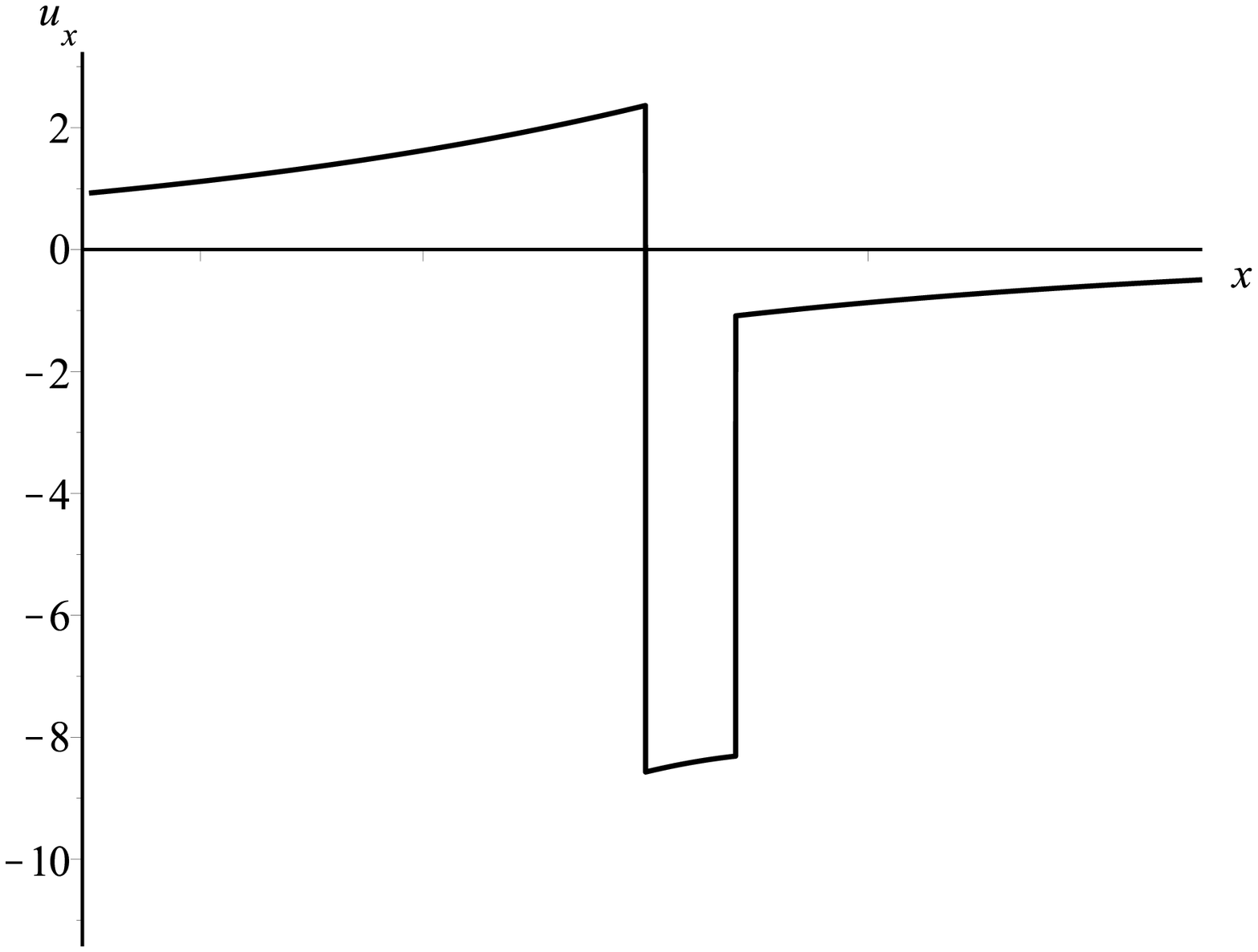}
\captionof{figure}{$t=0.5T$}
\end{subfigure}
\begin{subfigure}[t]{.3\textwidth}
\includegraphics[width=\textwidth]{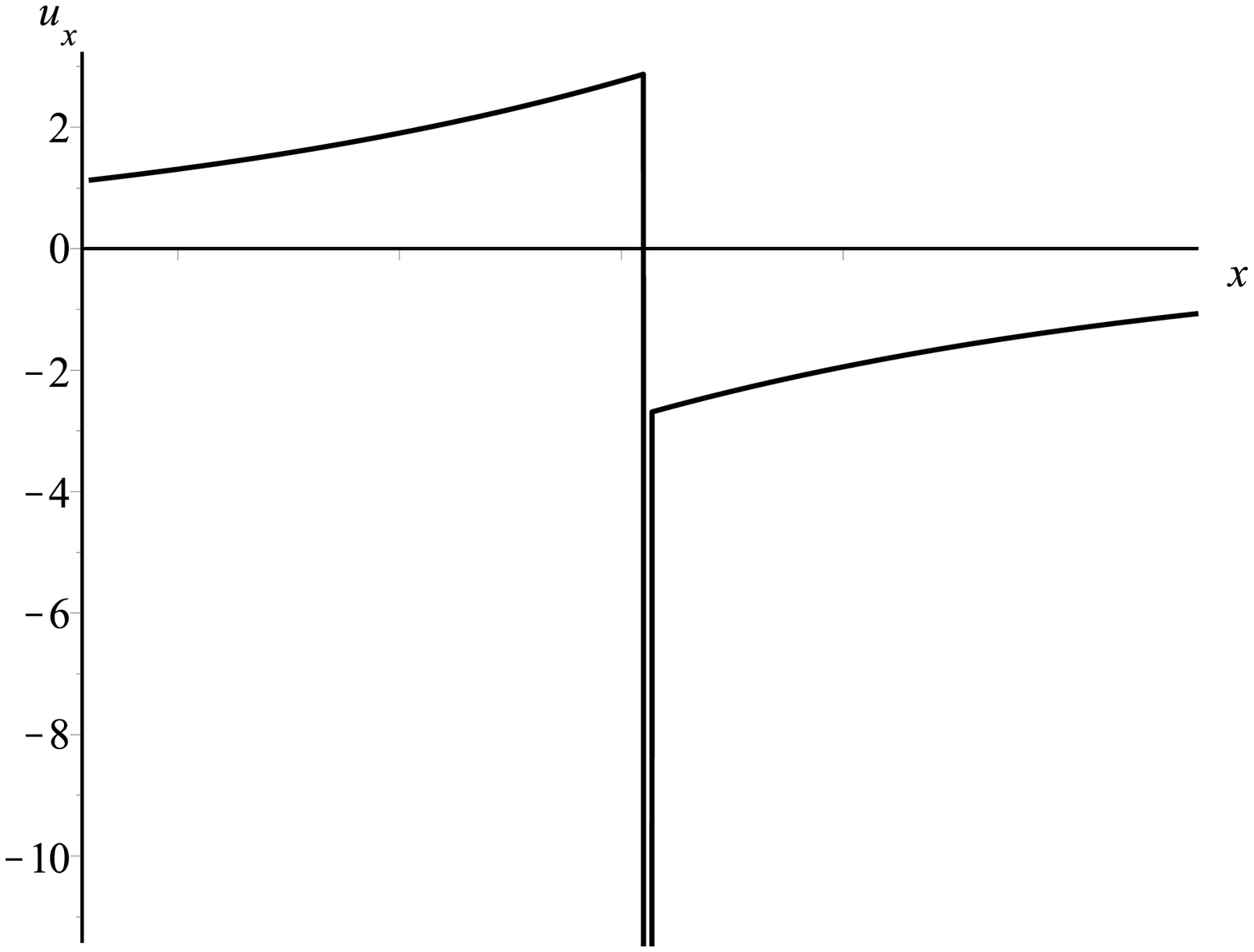}
\captionof{figure}{$t=0.72T$}
\end{subfigure}%
\begin{subfigure}[t]{.3\textwidth}
\includegraphics[width=\textwidth]{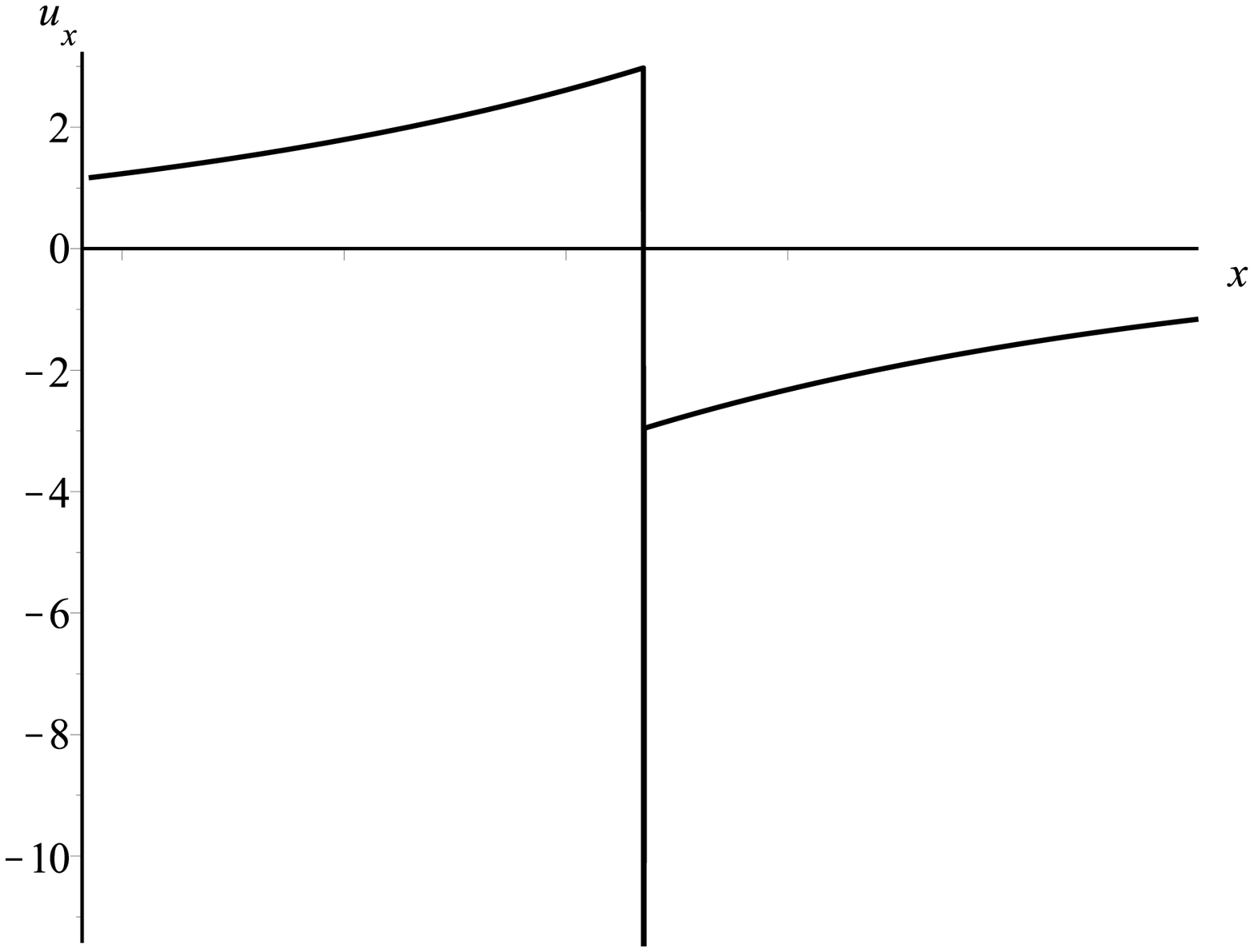}
\captionof{figure}{$t=0.9T$}
\end{subfigure}
\caption{Blow-up of slope in collision of peakon and anti-peakon in the case $p=3$}
\label{p=3-blowup}
\end{figure}

\begin{figure}[H]
\centering
\begin{subfigure}[t]{.3\textwidth}
\includegraphics[width=\textwidth]{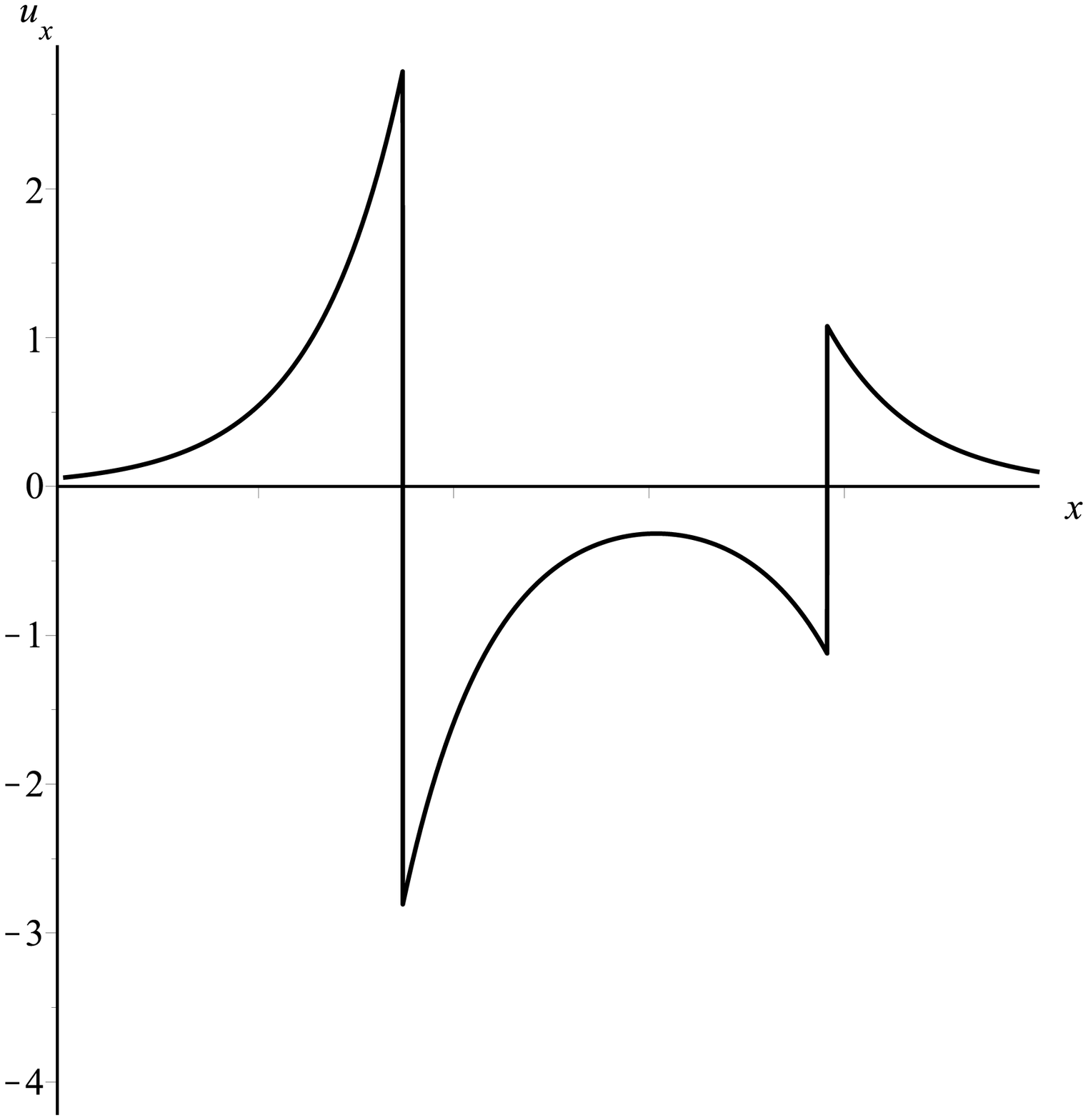}
\captionof{figure}{$t=0$}
\end{subfigure}%
\begin{subfigure}[t]{.3\textwidth}
\includegraphics[width=\textwidth]{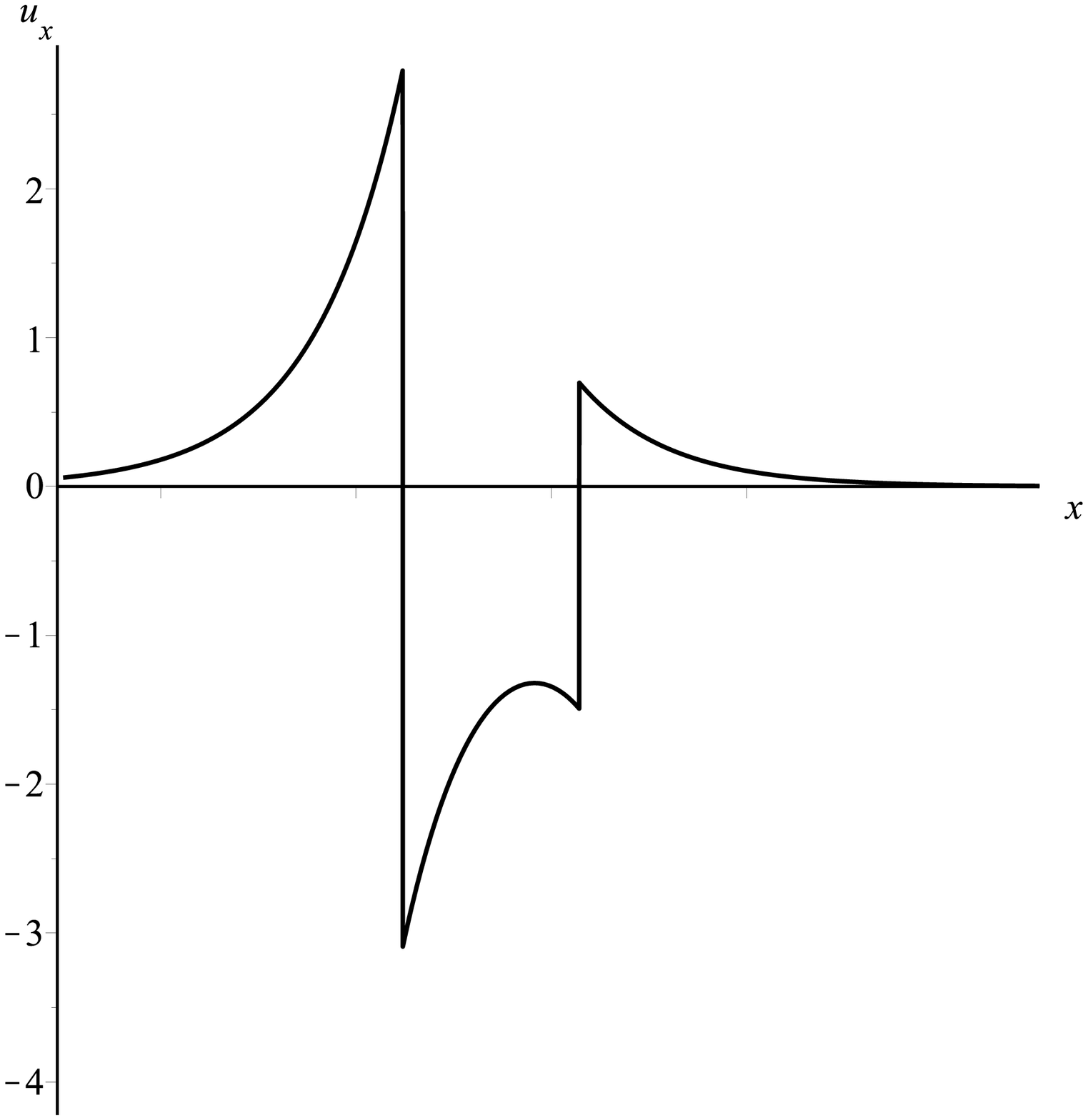}
\captionof{figure}{$t=0.5T$}
\end{subfigure}
\begin{subfigure}[t]{.3\textwidth}
\includegraphics[width=\textwidth]{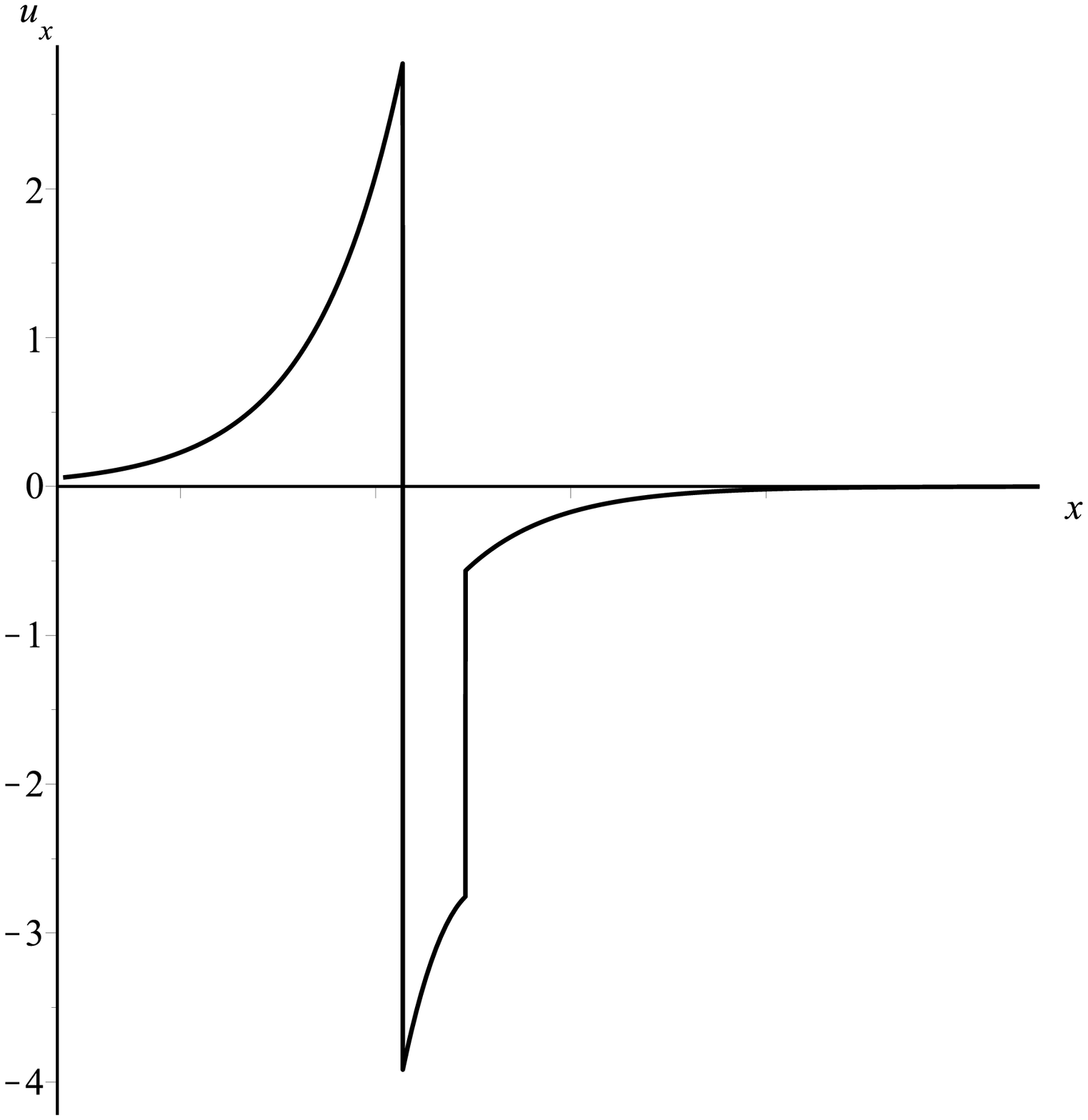}
\captionof{figure}{$t=0.72T$}
\end{subfigure}%
\begin{subfigure}[t]{.3\textwidth}
\includegraphics[width=\textwidth]{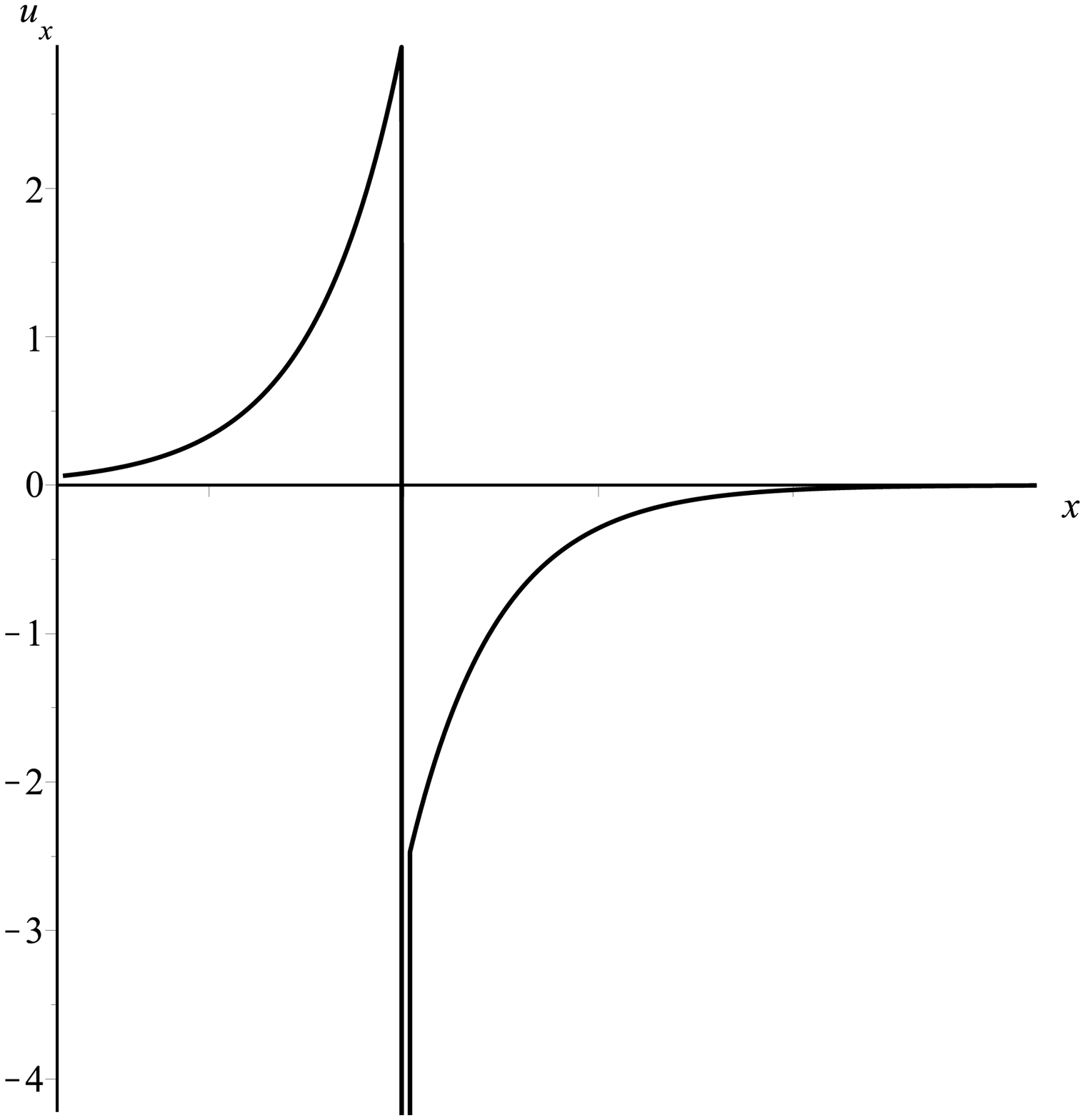}
\captionof{figure}{$t=0.9T$}
\end{subfigure}
\caption{Blow-up of slope in collision of peakon and anti-peakon in the case $p=4$}
\label{p=4-blowup}
\end{figure}

In all cases $p\neq 1$, 
the flow will evolve the initial amplitudes toward a stable attractor line. 
This evolution is shown in Figs.~\ref{p=3-collide} and~\ref{p=4-collide} 
for the cases $p=3$ and $p=4$, respectively, 
where the initial positions of the peakon and anti-peakon are chosen to be distinctly separated. 
We see that the peakon and anti-peakon collide such that their peak amplitudes 
become closer while the slope at locations $x$ in between the two peaks rapidly increases (without bound)
as relative separation between their positions decreases to zero in a finite time. 
This blow-up in the slope seen in Figs.~\ref{p=3-blowup} and~\ref{p=4-blowup} 
is an example of wave breaking. 

There is a qualitative explanation of why the blow-up in the slope $u_x$ 
between the two peaks in a collision solution \eqref{2peakonu} 
occurs in a finite time. 
Consider the asymptotic attractor solution 
$u(t,x) = \sqrt{H}\exp(-|x-\sqrt{H}^p t|)$
corresponding to the upper line \eqref{H-lines}. 
This solution arises from the initial condition 
$\alpha_1(0)=0$ and $\alpha_2(0)=\sqrt(H)$. 
The amplitude ODEs \eqref{2peakon-ampl1-eom}--\eqref{2peakon-ampl2-eom} yield 
\begin{equation}\label{attractor}
\alpha_1 = \frac{-\sqrt{H}}{\exp(2\sqrt{H}^p(T-t))-1},
\quad
\alpha_2 = \frac{\sqrt{H}\exp(2\sqrt{H}^p(T-t))}{\exp(2\sqrt{H}^p(T-t))-1}
\end{equation}
whereby $\alpha_1\rightarrow-\infty$ and $\alpha_2\rightarrow \infty$ 
as $t\rightarrow\infty$ such that $\alpha_1 + \alpha_2 = \sqrt{H}$ is constant
for all $t\geq 0$. 
Any solution having an initial condition close to $\alpha_1(0)=0$ and $\alpha_2(0)=\sqrt(H)$ 
will exhibit a similar long-time behaviour for $\alpha_1$ and $\alpha_2$,
as a consequence of continuous dependence of solutions on initial data for the ODEs \eqref{2peakon-ampl1-eom}--\eqref{2peakon-ampl2-eom}. 
Since $\alpha_1 + \alpha_2 \rightarrow \sqrt{H}$, 
the solution \eqref{2peakonu} remains continuous and bounded at all $x$ for $t\geq 0$, 
whereas the slope 
\begin{equation}\label{2peakonux}
u = \sgn(\beta_1-x)\alpha_1 e^{-|x-\beta_1|} +\sgn(\beta_2-x) \alpha_2 e^{-|x-\beta_2|} 
\end{equation}
has jump discontinuities at $x=\beta_1$ and $x=\beta_2$ 
and becomes unbounded at $x\rightarrow (\beta_1+\beta_2)/2$ 
(with $\beta_1-\beta_2\rightarrow 0$) as $t\rightarrow T<\infty$. 

The same kind of wave-breaking behaviour can be expected to occur in collisions 
between peakons and anti-peakons when $N>2$. 

\section{Concluding remarks}
\label{remarks}

At first sight, 
the gCHN equation \eqref{CHNm} seems closely analogous to the $b$-equation \eqref{CH-DP}: 
both equations unify two integrable equations, possess $N$-peakon solutions, 
and exhibit wave breaking phenomena. 
However, there are important differences. 
Firstly, 
the nonlinearities in the $b$-equation are purely quadratic,
whereas the gCHN equation has nonlinearities of degree $p+1$
and thereby connects two integrable equations with different nonlinearities. 
Secondly, 
the $H^1$ norm of solutions $u(t,x)$ is conserved 
for the $b$-equation only if $b=1$, 
when the $b$-equation reduces to the Camassa-Holm equation. 
In contrast, the $H^1$ norm is conserved for the gCHN equation for all $p\neq 0$.

In a subsequent work, 
we will explore further properties of the gCHN equation \eqref{CHNm}
and its multi-peakon solutions. 
There are numerous interesting questions. 
Can a wave-breaking result similar to those for the Camassa-Holm and Novikov
equations be established for classical solutions? 
How will the wave-breaking behaviour depend on $p$? 
In particular, a plausible criteria for wave-breaking is 
$\underset{t\rightarrow T}{\lim} \underset{x\in\Rnum}{\liminf}(u^{p-1}u_x)=-\infty$
which generalizes the criteria known \cite{ConEsc,JiaNi} 
in the Camassa-Holm case $p=1$ and the Novikov case $p=2$. 
In another direction, for any $p$ other than these two known integrable cases $p=1$ and $p=2$, 
does the equation have a Hamiltonian formulation 
or perhaps integrability properties?

\section*{Acknowledgements}

S.C. Anco is supported by an NSERC research grant. 
P.L. da Silva and I.L. Freire would like to thank FAPESP (scholarship n. 2012/22725-4 and grant n. 2014/05024-8) and CAPES for financial support. 
I.L. Freire is also partially supported by CNPq (grant n. 308941/2013-6). 
The referee is thanked for remarks which have improved this paper.

\appendix
\section{}

\subsection{Lie symmetries}
\label{liesymms}

To classify all of the Lie symmetries admitted by the 4-parameter equation \eqref{4parmCHN},
we first substitute a general coefficient function $P(t,x,u,u_t,u_x)$ 
into the symmetry determining equation \eqref{symmdeteq}. 
Next we eliminate $u_{txx}$, $u_{ttxx}$, $u_{txxx}$ 
through writing the equation in the solved form 
\begin{equation}\label{ut2x}
u_{txx} = u_t + au^p u_x -b u^{p-1} u_x u_{xx} -c u^p u_{xxx}
\end{equation}
and doing the same for its differential consequences. 
The determining equation \eqref{symmdeteq} then splits 
with respect to $u_{xx}$, $u_{tx}$, $u_{tt}$, $u_{xxx}$, $u_{ttx}$, $u_{xxxx}$ 
into a linear overdetermined system of 10 equations 
for $P(t,x,u,u_t,u_x),a,p,b,c$:
\begin{align}
& 
P_{u_xu_x}=0,
\quad
P_{u_tu_x}=0,
\quad
P_{u_tu_t}=0,
\label{symmeq1}\\
&
P_{uu_x}=0,
\quad
P_{uu_t}=0,
\quad
P_{xu_t}=0,
\label{symmeq2}\\
&
2P_{xu} + 2u_x P_{uu} + P_{xxu_x}=0, 
\label{symmeq3}\\
& 
cu^p( P_{tu_t} - P_{xu_x}) + cpu^{p-1}( u_tP_{u_t} +u_xP_{u_x} -P ) -P_{tu_x}=0,
\label{symmeq4}\\
& 
(p-1)bu^{p-2}u_x( u_tP_{u_t} +u_xP_{u_x} -P ) 
+ 3cu^p( P_{xu} +u_x P_{uu}) 
\nonumber\\&\quad
- bu^{p-1}(u_x(P_{u} - P_{tu_t}) + P_x)
-P_{tu}-u_tP_{uu} -2P_{xtu_x}=0,
\label{symmeq5}\\
& 
apu^{p-1} u_x( u_tP_{u_t} +u_xP_{u_x} -P ) 
+ (au^p u_x+u_t)( u_tP_{t} +2u_xP_{x} ) -au^p P_{x} 
\nonumber\\&\quad
+cu^p( u_x^{\,3}P_{uuu} + 3u_x^{\,2}P_{xuu} + 3u_xP_{xxu} +P_{xxx} )
+ bu^{p-1}( u_x^{\,3}P_{uu} + 2u_x^{\,2}P_{xu} + u_xP_{xx} ) 
\nonumber\\&\quad
+ u_x^{\,2}( P_{tuu} + u_tP_{uuu} ) + 2u_x( P_{xtu} + u_tP_{xuu} ) 
+u_tP_{xxu} + P_{xxt}-P_{t} =0.
\label{symmeq6}
\end{align}
Equation \eqref{symmeq1} shows that $P$ is a linear function of $u_t$ and $u_x$,
and hence 
\begin{equation}
P=\eta-\tau u_t-\xi u_x
\end{equation}
for some functions $\eta(t,x,u),\tau(t,x,u),\xi(t,x,u)$. 
After simplifying the remaining equations \eqref{symmeq2}--\eqref{symmeq6},
we obtain a system of 14 equations
\begin{align}
& 
\tau_x=0,
\quad
\tau_u=0,
\quad
\xi_u=0,
\quad
\eta_{xuu}=0,
\quad
\eta_{tuu}=0,
\label{pointsymmeq1}\\
&
\tau_{tu}-\eta_{uu}=0,
\quad
\xi_{xx}-2\eta_{xu}=0,
\quad
b\eta_{uu}+c\eta_{uuu}u=0,
\quad
4\xi_{x}-\xi_{xxx}=0,
\label{pointsymmeq2}\\
&
\xi_{t}+c(\xi_{x}-\tau_{t})u^p-cpu^{p-1}\eta=0,
\quad
\eta_{t}-\eta_{txx}+(a\eta_{x}-c\eta_{xxx})u^p=0,
\label{pointsymmeq3}\\
&
b\eta_{xx} -(a-c)((\xi_{x}+\tau_{t})u+p\eta)=0,
\label{pointsymmeq4}\\
&
4\xi_{tx}-2\eta_{tu}-2b\eta_{x}u^{p-1}+3c\xi_{xx}u^p=0,
\label{pointsymmeq5}\\
&
3\xi_{tu}-b(p-1)\eta u^{p-2}+b(\xi_{x}-\tau_{t}-\eta_{u})u^{p-1}-3\tau_{tu}u^p=0.
\label{pointsymmeq6}
\end{align}
We solve this linear overdetermined system by the following steps. 
First, an integrability analysis of 
the system of equations \eqref{pointsymmeq1}--\eqref{pointsymmeq6}
is carried out using the Maple package {\em rifsimp}, 
which yields 9 cases. 
Next, in each case the reduced system of equations is integrated 
using the Maple command {\em pdsolve}.  
Last, the solutions are merged, which leads to the following classification result. 

\begin{prop}\label{class-pointsymm}
(i) For any $p\neq 0$ and any $(a,b,c)\neq 0$, 
equation \eqref{4parmCHN} admits no contact symmetries.
(ii) The point symmetries admitted by equation \eqref{4parmCHN} 
for all $p\neq 0$ and all $(a,b,c)\neq 0$ 
consist of 
\begin{equation}\label{pointsymm-general}
\X_1=\parder{t},
\quad
\X_2=\parder{x} ,
\quad
\X_3= -pt\parder{t} +u\parder{u} . 
\end{equation}
(iii) Equation \eqref{4parmCHN} admits additional point symmetries 
only in the following cases:
\begin{align}
{\rm (a)}\qquad
\label{pointsymm-a}
& \X_{4\rm a}=at\parder{x} +\parder{u} 
\quad\text{ iff }\quad
p=1,
\quad
a=c
\\\nonumber\\
{\rm (b)}\qquad
\label{pointsymm-b}
& X_{4\rm b}=\exp(\pm 2x)( \pm \parder{x} + u\parder{u} ) 
\quad\text{ iff }\quad
p=2, 
\quad
a=4c,
\quad
b=3c
\end{align}
\end{prop}

The symmetries $X_1$, $X_2$, $X_3$ respectively generate 
one-dimensional point transformation groups consisting of 
time-translations $t\rightarrow t+\epsilon$, 
space-translations $x\rightarrow x+\epsilon$, 
and scalings $t\rightarrow \exp(-b\epsilon)t$, $u\rightarrow \exp(\epsilon)u$, 
with group parameter $\epsilon\in\Rnum$. 
The extra symmetry $\X_{4\rm a}$ generates 
the one-dimensional point transformation group 
$u\rightarrow u+\epsilon$, $x\rightarrow x+\epsilon at$,
which is a Galilean boost,
and the other extra symmetry $\X_{4\rm b}$ generates 
the one-dimensional point transformation group 
$u\rightarrow (1-2\epsilon\exp(\pm2x))^{1/2}u$, 
$x\rightarrow x\pm\tfrac{1}{2}\ln(1-2\epsilon\exp(\pm2x))$,
which is a non-rigid dilation. 

A symmetry analysis of particular equations in the 4-parameter family \eqref{4parmCHN}
can be found in Refs.~\cite{BozFreIbr,SilFre15,ClaManPri}. 

\subsection{Low-order multipliers}

To classify all 1st-order multipliers admitted by the 4-parameter equation \eqref{4parmCHN},
we first substitute the expression \eqref{Q1storder} 
into the determining equation \eqref{Qdeteq},
which splits into a linear overdetermined system of 5 equations 
for $Q(t,x,u,u_t,u_x),p,a,b,c$. 
The system contains the equations 
\begin{equation}
Q_{u_t}=0,
\quad
Q_{u_x}=0
\end{equation}
which yield
\begin{equation}
Q=Q_0(t,x,u) . 
\end{equation}
After the remaining 3 equations are split with respect to $u_t$ and $u_x$, 
we obtain the following system of 8 equations
\begin{align}
&
Q_{0xu}=0,
\quad
Q_{0uu}=0,
\label{Qeq1}\\
&
(p-1)(3pc-2b)Q_{0x}=0,
\quad
p((p+1)c-b)Q_{0u}=0,
\quad
(3pc-b)Q_{0xx}=0,
\label{Qeq2}\\
&
(p-1)(pc-b)Q_{0}+(2pc-b)Q_{0u}u=0,
\label{Qeq3}\\
&
Q_{0tu}+(3pc-b)Q_{0x}u^{p-1}=0,
\label{Qeq4}\\
&
Q_{0xxt} -Q_{0t} +(cQ_{0xxx} -aQ_{0x})u^p=0.
\label{Qeq5}
\end{align}
We solve this linear overdetermined system by the same three steps
used in solving the symmetry system \eqref{pointsymmeq1}--\eqref{pointsymmeq6}.
This yields the five distinct cases
presented in parts (i) and (ii) of Proposition~\ref{class-loworderQ}. 

Finally, by splitting and simplifying the determining equation \eqref{Qdeteq} 
for second-order multipliers of the form \eqref{Qfunctm}, 
we obtain a linear overdetermined system of 13 equations for 
$Q(u,u_x,u_{xx}),a,p,b,c$. 
One of the equations in this system is given by 
\begin{equation}
Q_{u_x}=0
\end{equation}
which yields
\begin{equation}
Q=Q_0(u,u_{xx}) . 
\end{equation}
The remaining 10 equations then split with respect to $u_x$, 
leading to a system of 6 equations
\begin{align}
&
Q_{0uu}=0,
\quad
Q_{0uu_{xx}}=0,
\quad
Q_{0u_{xx}u_{xx}}=0,
\label{2ndQeq1}\\
&
(p-1)(pc-2b)Q_{0u_{xx}}=0,
\label{2ndQeq2}\\
&
p(aQ_{0u_{xx}}+((p+1)c-b)Q_{0u}) =0
\label{2ndQeq3}\\
&
(p-1)((pc-b)Q_{0} -u_{xx}Q_{0u_{xx}} - (pc-b)Q_{0u}u) =0.
\label{2ndQeq4}
\end{align}
Solving this linear overdetermined system by the same steps
used in solving the multiplier system \eqref{Qeq1}--\eqref{Qeq5},
we obtain the two distinct cases
presented in part (iii) of Proposition~\ref{class-loworderQ}.

\end{document}